\documentclass[letter]{aa}  
\usepackage{graphicx}
\usepackage{txfonts}

\newcommand{\mlp}[1]{{#1}}

\newcommand{\stefd}[2]{\protect{\bgroup\markoverwith{\textcolor{red}{\rule[0.5ex]{2pt}{0.4pt}}} \ULon{#1}} \textcolor{red} {#2}}
\newcommand{\mlpd}[2]{\protect{\bgroup\markoverwith{\rule[0.5ex]{2pt}{0.4pt}} \ULon{#1}} {\bf #2}}
\newcommand{\alexd}[2]{\protect{\bgroup\markoverwith{\textcolor{green}{\rule[0.5ex]{2pt}{0.4pt}}} \ULon{#1}} \textcolor{green} {#2}}

\def\hdu18{HD\,189733\,b}
\def\hd20{HD\,209458\,b}
\def\gj34{GJ\,3470\,b}

\def\co2{CO$_2$}
\def\ch4{CH$_4$}
\def\h2{H$_{2}$}
\def\h2o{H$_2$O}

\def\hes{He(1$^{1}$S)}
\def\het{He(2$^{3}$S)}
\def\mlr{$\dot M$}
\def\lya{Ly$\alpha$}

\def\rp{$R_{\rm P}$}

\begin{document} 

\title{Evidence of energy-, recombination-, and photon-limited escape regimes in giant planet H/He atmospheres}
%
%
\titlerunning{Evidence of energy-, recombination-, and photon-limited escape regimes}
\author{
M.~Lamp{\'o}n\inst{1}, 
M.~L\'opez-Puertas\inst{1},
S.~Czesla\inst{2},
A.~S\'anchez-L\'opez\inst{3},
L.\,M.~Lara\inst{1},
M.~Salz\inst{2},
J.~Sanz-Forcada\inst{4},
K.~Molaverdikhani\inst{5,6},
A.~Quirrenbach\inst{6},
E.~Pall{\'e}\inst{7,8},
J.\,A.~Caballero\inst{4},
Th.~Henning\inst{5},
L.~Nortmann\inst{9},
P.\,J.~Amado\inst{1},
D.~Montes\inst{10},
A.~Reiners\inst{9}, and 
I.~Ribas\inst{11,12}
}

\institute{Instituto de Astrof{\'i}sica de Andaluc{\'i}a (IAA-CSIC), Glorieta de la Astronom{\'i}a s/n, 18008 Granada, Spain\\
\email{mlampon@iaa.es}
\and
Hamburger Sternwarte, Universit{\"a}t Hamburg, Gojenbergsweg 112, 21029 Hamburg, Germany
\and
Leiden Observatory, Leiden University, Postbus 9513, 2300 RA, Leiden, The Netherlands
\and
Centro de Astrobiolog{\'i}a (CSIC-INTA), ESAC, Camino bajo del castillo s/n, 28692 Villanueva de la Ca{\~n}ada, Madrid, Spain
\and
Max-Planck-Institut f{\"u}r Astronomie, K{\"o}nigstuhl 17, 69117 Heidelberg, Germany
\and
Landessternwarte, Zentrum f\"ur Astronomie der Universit\"at Heidelberg, K\"onigstuhl 12, 69117 Heidelberg, Germany
\and
Instituto de Astrof{\'i}sica de Canarias (IAC), Calle V{\'i}a L{\'a}ctea s/n, 38200 La Laguna, Tenerife, Spain
\and
Departamento de Astrof{\'i}sica, Universidad de La Laguna, 38026  La Laguna, Tenerife, Spain
\and
Institut f{\"u}r Astrophysik, Georg-August-Universit{\"a}t, Friedrich-Hund-Platz 1, 37077 G{\"o}ttingen, Germany
\and
Departamento de F{\'i}sica de la Tierra y Astrof{\'i}sica \& IPARCOS-UCM (Instituto de F{\'i}sica de Part{\'i}culas y del Cosmos de la UCM), Facultad de Ciencias F{\'i}sicas, Universidad Complutense de Madrid,  28040 Madrid, Spain
\and
Institut de Ci\`encies de l'Espai (CSIC-IEEC), Campus UAB, c/ de Can Magrans s/n, 08193 Bellaterra, Barcelona, Spain
\and
Institut d'Estudis Espacials de Catalunya (IEEC), 08034 Barcelona, Spain
}

\authorrunning{M.Lamp\'on et al.}
\date{Received 26 January 2021 / Accepted 19 March 2021}

\abstract
{Hydrodynamic escape is the most efficient atmospheric mechanism of  planetary mass loss and has a large impact on planetary evolution.
Three hydrodynamic escape regimes have been identified theoretically:
energy-limited, recombination-limited, and photon-limited. However, no evidence of these regimes had been reported until now. 
Here, we report  evidence of these three regimes 
via an analysis of a helium~{\sc i} triplet at 10830\,\AA\ and \lya\ absorption involving a 1D hydrodynamic model that allows us to estimate hydrogen recombination and advection rates. 
In particular, we show that \hd20 is in the energy-limited regime, \hdu18 is in the recombination-limited regime, and \gj34 is in the photon-limited regime. These exoplanets can be considered as benchmark cases for their respective regimes.
}

\keywords{planets and satellites: atmospheres -- planets and satellites: gaseous planets -- planets and satellites: individual: \hd20 -- planets and satellites: individual: \hdu18 -- planets and satellites: individual: \gj34 }
\maketitle
%

\section{Introduction}
\label{Intro}

Understanding the diversity, evolution, and formation of planets requires a profound knowledge of their atmospheric escape. This statement especially applies to planets that, at some stage in their history, 
experience the most effective atmospheric process of mass loss, 
the hydrodynamic atmospheric escape  (also known as blow-off escape). A planetary atmosphere is in hydrodynamic escape when the absorbed stellar high-energy flux is strong enough to produce substantial heating of the atmosphere, which triggers a massive gas outflow, driving the escaping gas to supersonic bulk velocities. 
This process can lead from a loss of a fraction of the original atmosphere to its complete removal \citep[see e.g.][]{Baraffe_2004,Baraffe_2005,Garcia_munoz_2019} and hence can shape the current planetary population \citep[e.g. forming the sub-Jovian desert and the evaporation valley;][]{Owen_2013,Lopez_2013}.
Hydrodynamic escape can also significantly change the composition of the primordial atmospheres 
and thus affects the planetary mass-radius relationship \cite[see e.g.][]{Hu_2015,Malsky_2020}.
 
Several models have been developed for studying planetary atmospheres undergoing hydrodynamic escape \cite[e.g.][]{Watson_1981,Lammer_2003,Yelle_2004,Tian2005,Garcia_munoz_2007,Murray_Clay_2009,Owen_2016,Salz2016,Bourrier_2013,Shaikhislamov_2020}. 
In particular, \cite{Murray_Clay_2009} studied atmospheric mass loss as a function of the incident ultraviolet (UV) flux for a typical hot Jupiter. They found that, under high irradiation, the mass-loss rate was not limited by the radiative energy deposited in the atmosphere, but by the radiation-recombination equilibrium of the escaping gas. 
When the mass-loss rate is limited by the energy deposition, it scales linearly with the incident UV flux, which is efficiently converted into the work that lifts the gas. When the mass-loss rate is limited by the radiation-recombination equilibrium, it approximately scales to the square root of the  incident UV flux since a large fraction of the absorbed energy is lost via radiative cooling.
\cite{Owen_2016} found that the mass-loss rate of moderately irradiated low gravity planets can be limited by the number of incoming ionising photons, instead of the absorbed energy or the radiation-recombination equilibrium.
In this limit, the mass-loss rate scales to the square of the effective absorption radius of the planet.
Therefore, it is theoretically possible to identify three distinct hydrodynamic atmospheric escape regimes in hydrogen-dominated atmospheres: energy-limited, recombination-limited, and photon-limited. 

This classification is important when using the energy-limited approximation for estimating the planetary mass-loss rate
\cite[see e.g.][]{Murray_Clay_2009,Bear_2011,Owen_2016,Salz_2015,Wang_2018}.
This approach, used together with an average heating efficiency of 10--20\% \citep[the most probable range in hydrogen-dominated atmospheres;][]{Shematovich_2014}, largely overestimates the mass-loss rate of planets that are in the recombination-limited regime. 
It might also lead to significantly inaccurate results if applied to atmospheres in the photon-limited regime as the effective absorption radius could differ considerably from the planetary radius.
Hence, the applicability of this approach largely depends on the hydrodynamic escape regime.

Despite the central role of hydrodynamic escape regimes, a lack of suitable observations has prevented their confirmation. 
While \lya\ absorption measurements are promising tracers of this process, they can only be carried out from space and are adversely affected by interstellar absorption and geocoronal emission. This is not the case for the recent observations of the He~{\sc i} 2$^3$S--2$^3$P triplet transitions\footnote{At wavelengths\,10832.06, 10833.22, and 10833.31\,\AA\ in a vacuum; it is often referred to as their air wavelengths of  10830\,\AA.}, hereafter \het\ transitions; they provide new opportunities to explore the hydrodynamic atmospheric escape because these lines can be observed with ground-based telescopes and they are not affected by interstellar absorption \cite[e.g.][]{Spake_2018,Nortmann2018,Allart2018}. \mlp{Some examples of how these measurements help to better characterise the escaping atmospheres of gas giants were provided by  
\cite{Allart2018}, \cite{Mansfield_2018}, \cite{Lampon2020a,Lampon2020b}, and 
\cite{Shaikhislamov_2020}.}

In this work we study the hydrodynamic escape regimes of the giant planets \hd20, \hdu18, and \gj34.
The hot Jupiters \hd20 and \hdu18 undergo hydrodynamic escape, as shown by the detection of heavy species escaping in large amounts, such as O\,{\sc i} and C\,{\sc ii} in \hd20 as reported by \cite{VidalMadjar2004} and O\,{\sc i} in \hdu18 as documented by \cite{Ben_Jaffel_2013}. \gj34 also undergoes hydrodynamic escape as the outflow bulk radial velocities derived from the \het\  measurements far exceed the sound speed \cite[see Sect. 4.1.3 of][]{Lampon2020b}. These planets have different bulk parameters and stellar irradiation levels, which makes studying their escape regimes very interesting.

The letter is organised as follows. In Sect.\,\ref{sec:criteria} we discuss the three regimes that have been theoretically proposed as well as the parameters required to distinguish between them: the ratio of recombination to advection rates and, equivalently, the heating efficiency.  
In Sect.\,\ref{modelling_results} we summarise the  previous modelling and analysis of the three planets on which the calculations of these parameters have been based. In Sect.\,\ref{analysis} we present the results and discussion, and in Sect.\,\ref{summary} we provide a summary and our main conclusions.

\section{Criteria to differentiate the hydrodynamic atmospheric escape regimes}
\label{sec:criteria}

Physically, the three regimes can be distinguished by the production and losses of neutral H \cite[see][]{Murray_Clay_2009,Bear_2011,Owen_2016} 
as well as by the conversion of the absorbed stellar radiative energy into the work that drives the escape
\cite[see ][]{Murray_Clay_2009,Salz_2015}.

\subsection{Ratio of recombination to advection rates}
\label{sec:criteria_ratio}

The steady-state continuity equation of neutral H species, H${^0}$, can be expressed as (see Appendix\,\ref{ap:derivation})
\begin{equation}
    -v_{\rm r} \frac{\partial f_{H^0}}{\partial r} + f_{H^+}\ n_{H^+}\ \alpha_{H} - f_{H^0}\ J_H = 0 ,
    \label{eq:H}
\end{equation}
where $r$ is the distance from the centre of the planet; $v_{\rm r}$ is the bulk radial velocity of the gas; $f_{\rm {H^0}}$ and $f_{\rm {H^+}}$ are the mole fractions of H${^0}$ and the ionised H, H${^+}$, respectively; n$_{\rm H^+}$ is the H${^+}$ number density; and $\alpha_{\rm H}$ and $J_{\rm H}$ are the H recombination and photo-ionisation rates. 
We define the recombination production rate as  P$_{\textrm{rec}}$\,=\,$f_{\rm {H^+}}\,n_{\rm H^+}\,\alpha_{\rm H}$ and the advection rate as P$_{\textrm {adv}}$\,=\,$-v_{\rm{r}}\,\partial f_{\rm {H^0}}/\partial\,r$.

Similarly, and like \cite{Owen_2016}, we consider the ionisation front (IF) as the region where the atmosphere is partially ionised, that is, from 
the altitude where $f_{\rm {H^0}}$ is close to one -- $f_{\rm {H^0}}$\,$\geq$\,0.95, where the optical depth of the X-ray and extreme UV (XUV) radiation, $\tau _{\rm XUV}$, is $\geq$\,1 -- up to the altitude where the atmosphere is almost fully ionised -- $f_{\rm {H^0}}$\,$\leq$\,0.05 and $\tau_{\rm XUV}$\,$\ll 1$. In this region, the photo-ionisation rate is greater than the recombination rate and advection is non-negligible.

Following \cite{Murray_Clay_2009} and \cite{Owen_2016}, we characterise the different hydrodynamic atmospheric escape regimes as follows: (1) In the recombination-limited regime, the IF region is very narrow compared with the scale of the flow. The gas is almost in radiation-recombination equilibrium (i.e. advection is negligible,  $P_{\rm rec}/P_{\rm adv}$\,$\gg$\,1) in practically all of the upper atmosphere. 
(2) In the photon-limited regime, the photo-ionisation rate is much faster than the recombination rate such that the recombination process is negligible in the whole upper atmosphere and $P_{\rm rec}/P_{\rm adv}$\,$\ll$1. Consequently, the IF region occupies the whole upper atmosphere of the planet.
(3) The energy-limited regime is an intermediate case in between the recombination- and the photon-limited regimes. The IF is wide, although it does not encompass the entire flow as in the photon-limited regime. 
The advection and the recombination rates are non-negligible in the whole upper atmosphere.

\subsection{Heating efficiency criterion}
\label{sec:criteria_heating}
Although the ratio of recombination to advection rates provides us with a direct way of determining the hydrodynamic escape regime, it is also useful to look at these ratios from the point of view of heating efficiency, as previously done by several authors. This has the advantage of providing us with better insight into the physical processes behind the escape.
The heating efficiency, $\eta$, is usually defined as the fraction of the absorbed stellar radiative energy, {\textrm{W$_{\textrm{hv}}$}},
which is converted into the kinetic energy of the gas, namely, 
\begin{align}
    \eta = \frac{\textrm{W}_{\textrm{hv}} - {\textrm{W}_\textrm{e}} - \textrm{W}_{\textrm{cool}}}{\textrm{W}_{\textrm{hv}}}, 
    \label{eq:heating}
\end{align}
where W$_{\textrm e}$ is the rate of energy lost by photo-ionisation and photo-electron impact processes
\cite[as the excitation and ionisation of atoms by collisions with photoelectrons; see e.g.][]{Shematovich_2014} and  W$_{\textrm{cool}}$ is the radiative cooling rate, mostly produced by \lya\, and free-free emission \citep{Salz_2015}. Thus, W$_{\textrm{hv}}\,-$\,W$_{\textrm{e}}\,-$ W$_{\textrm{cool}}$ represents the energy from the absorbed stellar radiation that is converted into the kinetic energy of the gas species (atoms, ions, and electrons), the so-called thermal heating.

In the energy- and photon-limited regimes, radiative cooling is moderate or negligible, while in the recombination-limited regime it is a considerable fraction of the absorbed energy. 
If radiative cooling is neglected, the heating efficiency of stellar XUV radiation is about 0.1--0.2 in hydrogen-dominated planetary upper atmospheres \cite[see e.g.][]{Shematovich_2014,Salz_2015}. 
Consequently, the heating efficiency of the energy-limited and the photon-limited regimes is expected to be in the range of 0.1--0.2. In contrast, the heating efficiency of the recombination-limited regime is expected to be much less than 0.1 \cite[see e.g.][]{Salz_2015}.

\section{Recombination and advection rates and heating efficiencies} \label{modelling_results}

The recombination and advection rates of H$^0$ and the heating efficiencies analysed here are based on the modelling of the observations of the \het\ absorption of \hd20, \hdu18, and \gj34 that were performed by \cite{Lampon2020a,Lampon2020b}.
They used a 1D hydrodynamic model with a non-local thermodynamic equilibrium treatment for the \het\ state, 
and, by comparing the simulated and observed \het\ absorption, they derived well-constrained relationships \mlp{and narrow intervals} between the mass-loss rate, $\dot M$, and the temperature, $T$, in the upper atmosphere of these exoplanets.
Additionally, by comparing the H$^0$ density distribution with those derived from \lya\ measurements,
they constrained the H/He mole-fraction ratio.

\begin{figure}[htbp]
\includegraphics[angle=0, width=\columnwidth]{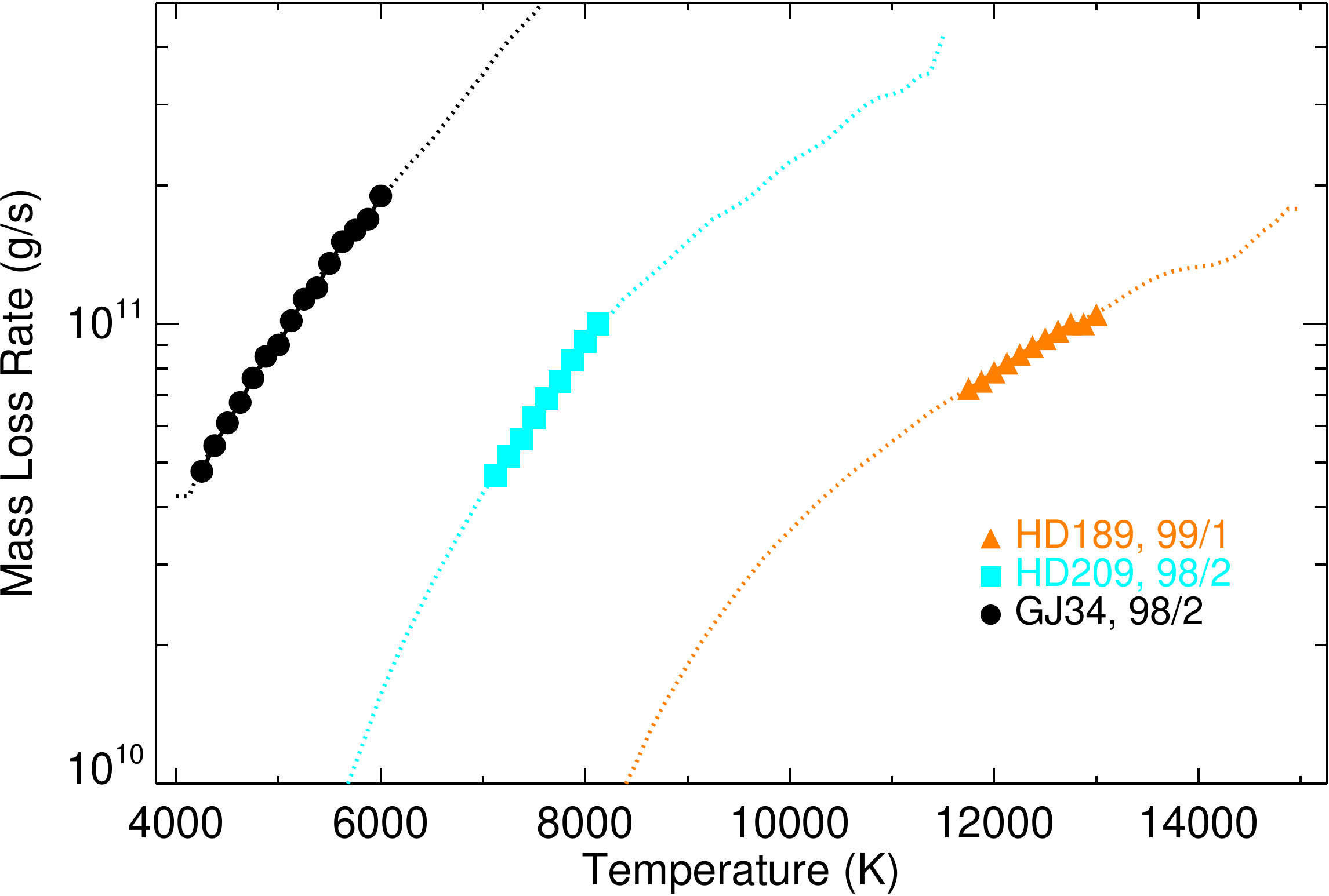}
\caption{Ranges of temperature and mass-loss rates for HD 209458\,b, HD\,189733\,b, and GJ\,3470\,b for H/He ratios of 98/2, 99/1, and 98/2, respectively, derived by \citet{Lampon2020a,Lampon2020b}. Symbols show the derived ranges and dotted lines the extended ranges (see Sect. \ref{sec:ionization}).}
\label{chi2} 
\end{figure}

 The \het\ production is controlled by electron concentration via recombination with He$^+$ as well as by electron collision excitation of \hes\ \cite[see][]{Oklopcic2018,Lampon2020a}. As electrons are predominantly produced by the photo-ionisation of H and lost by advection and H$^+$ recombination, the \het\ concentration profile is thus mainly controlled by the H production and loss rates. Therefore, by constraining the \het\ concentration profile, \mlp{as was done by \cite{Lampon2020a,Lampon2020b},} we simultaneously constrain the production and loss rates of H.
\mlp{Hence, in this work we analysed the production and loss rates of H corresponding to the $T$-$\dot M$ ranges and H/He ratios of 99/1 for \hdu18 and 98/2 for \hd20 and \gj34, which were previously obtained by \cite{Lampon2020a,Lampon2020b} and are summarised in Fig.\,\ref{chi2}.}

\begin{figure}[htbp]
\includegraphics[angle=0, width=\columnwidth]{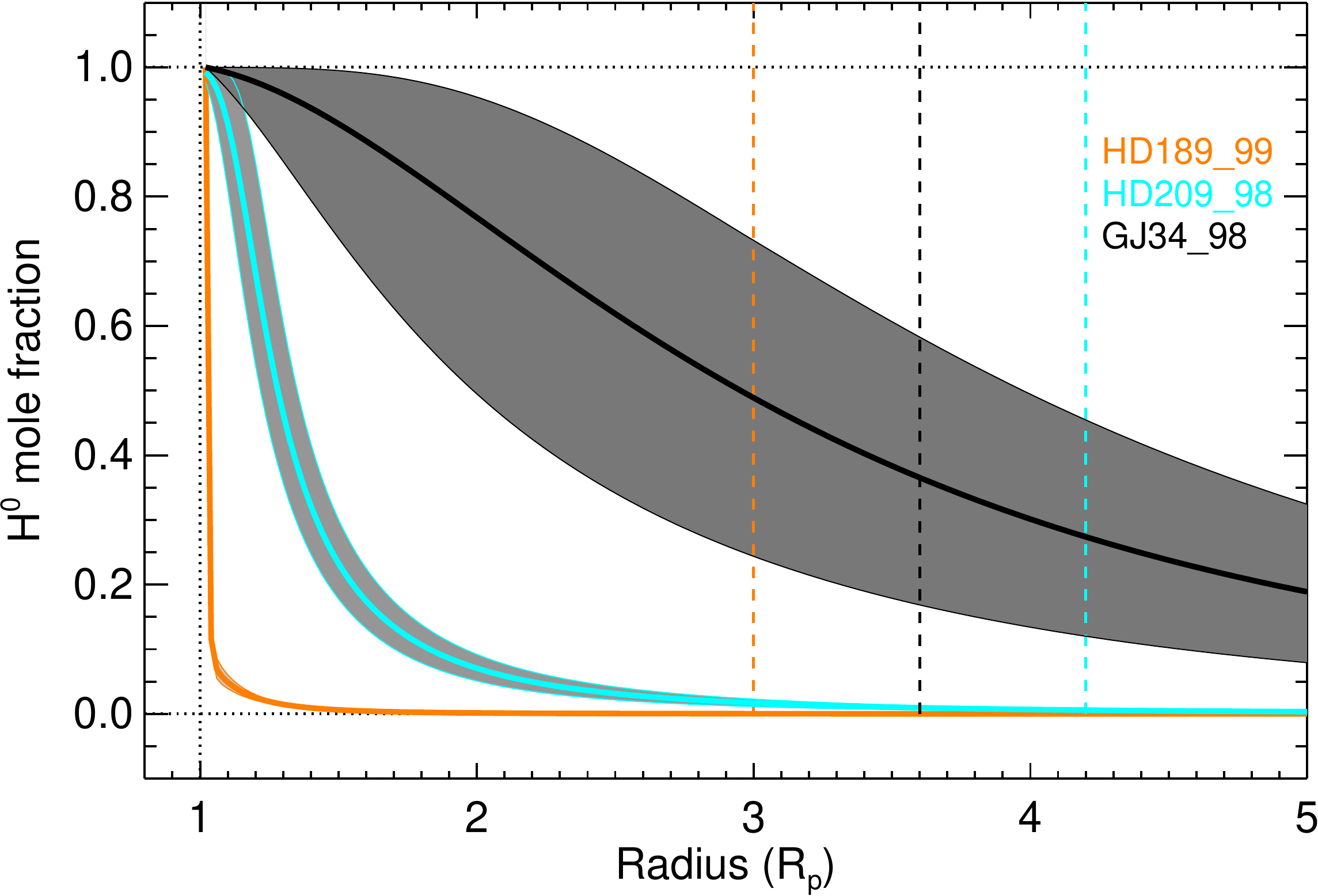}
\caption{Neutral H mole fraction profiles resulting from the measured \het\ and \lya\ absorption spectra for \hdu18, \hd20, and \gj34 with the representative H/He ratios of 99/1, 98/2, and 98/2, respectively.  The grey shaded areas correspond to the temperature and mass-loss rate ranges (symbols) in Fig.\,\ref{chi2}. The thicker solid lines are the mean profiles. The vertical dashed lines indicate the mean Roche lobes.} 
\label{hmol} 
\end{figure}

\begin{figure*}
\includegraphics[angle=0, width=0.66\columnwidth]{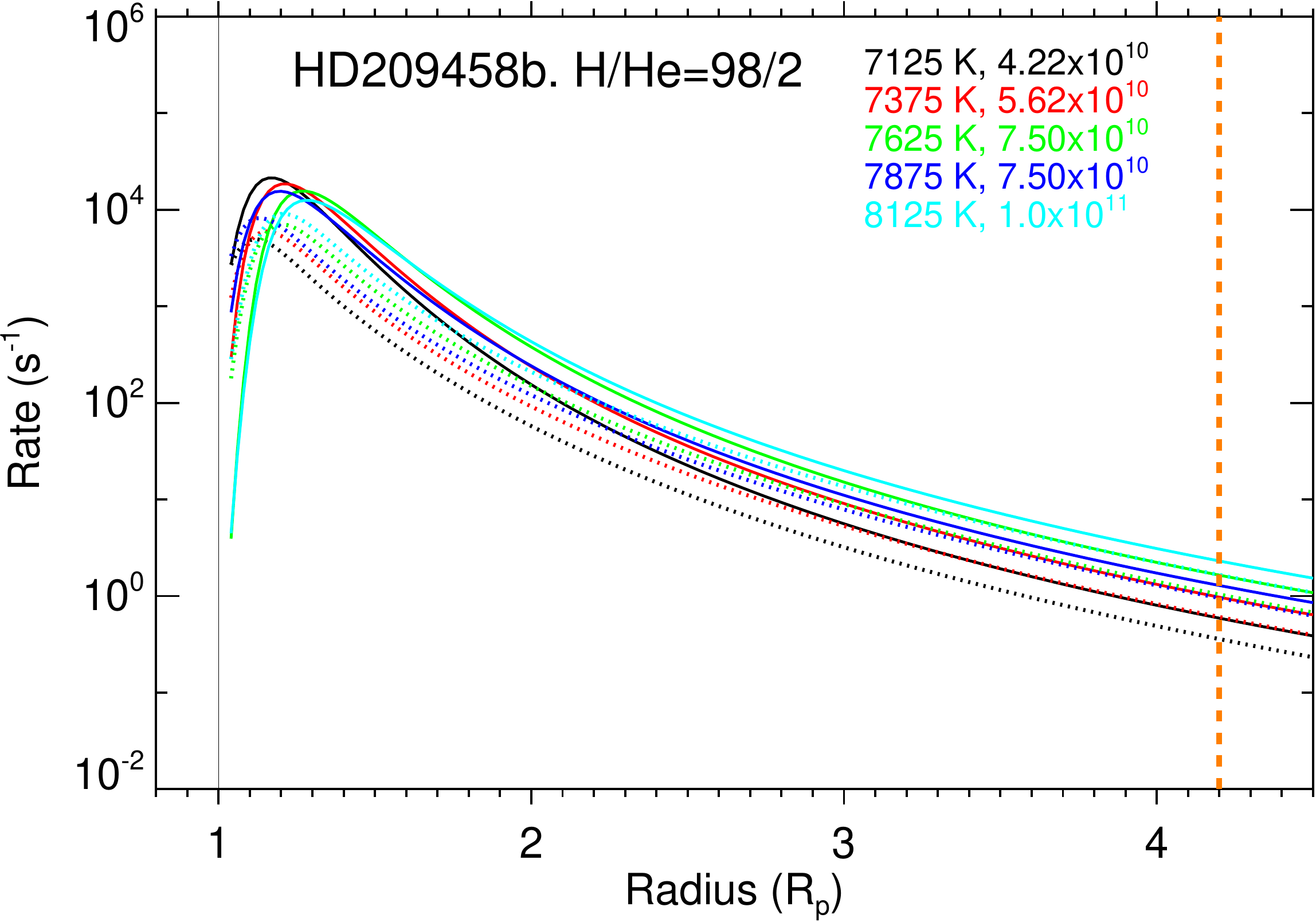}
\includegraphics[angle=0, width=0.66\columnwidth]{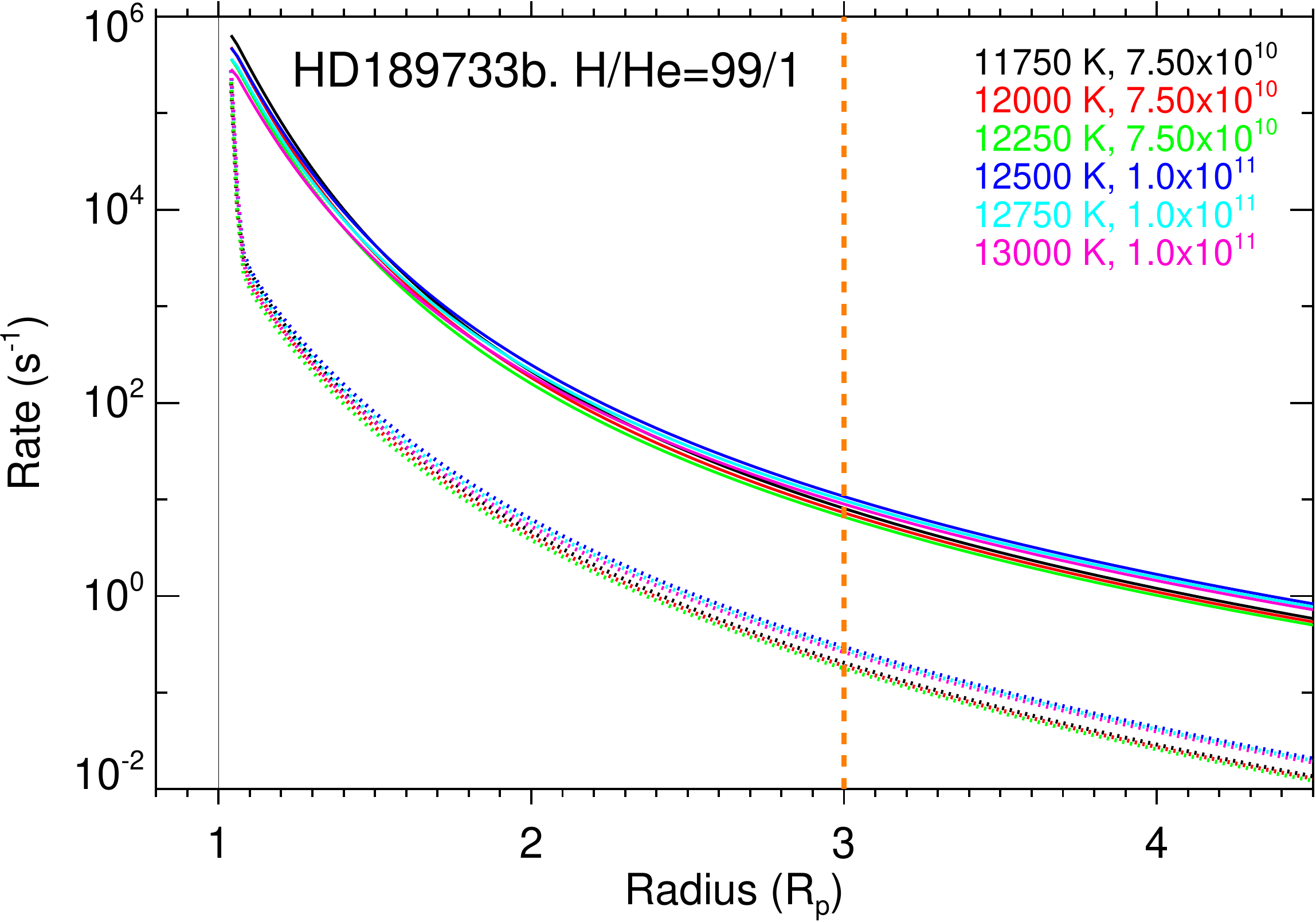}
\includegraphics[angle=0, width=0.66\columnwidth]{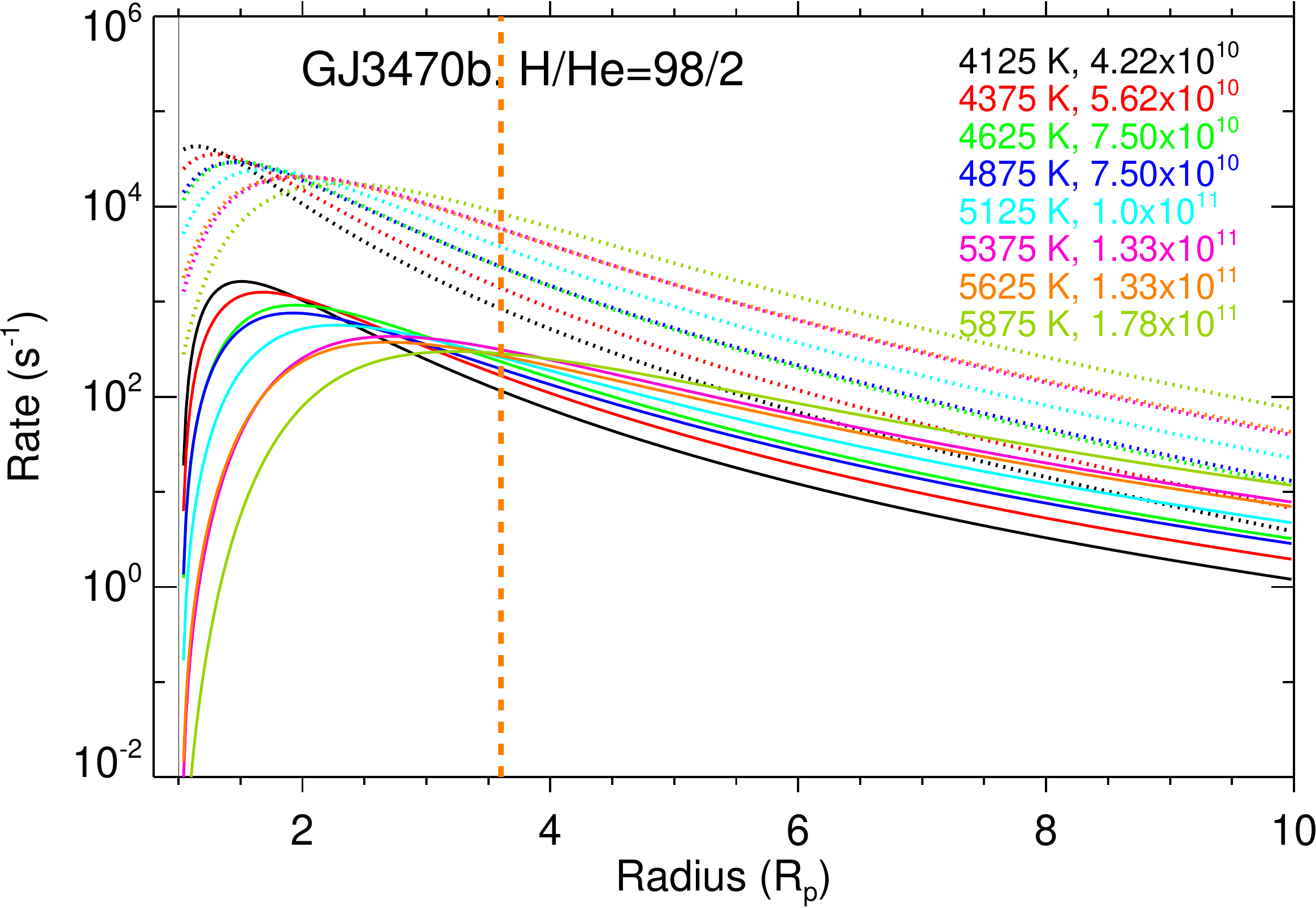}

\includegraphics[angle=0, width=0.66\columnwidth]{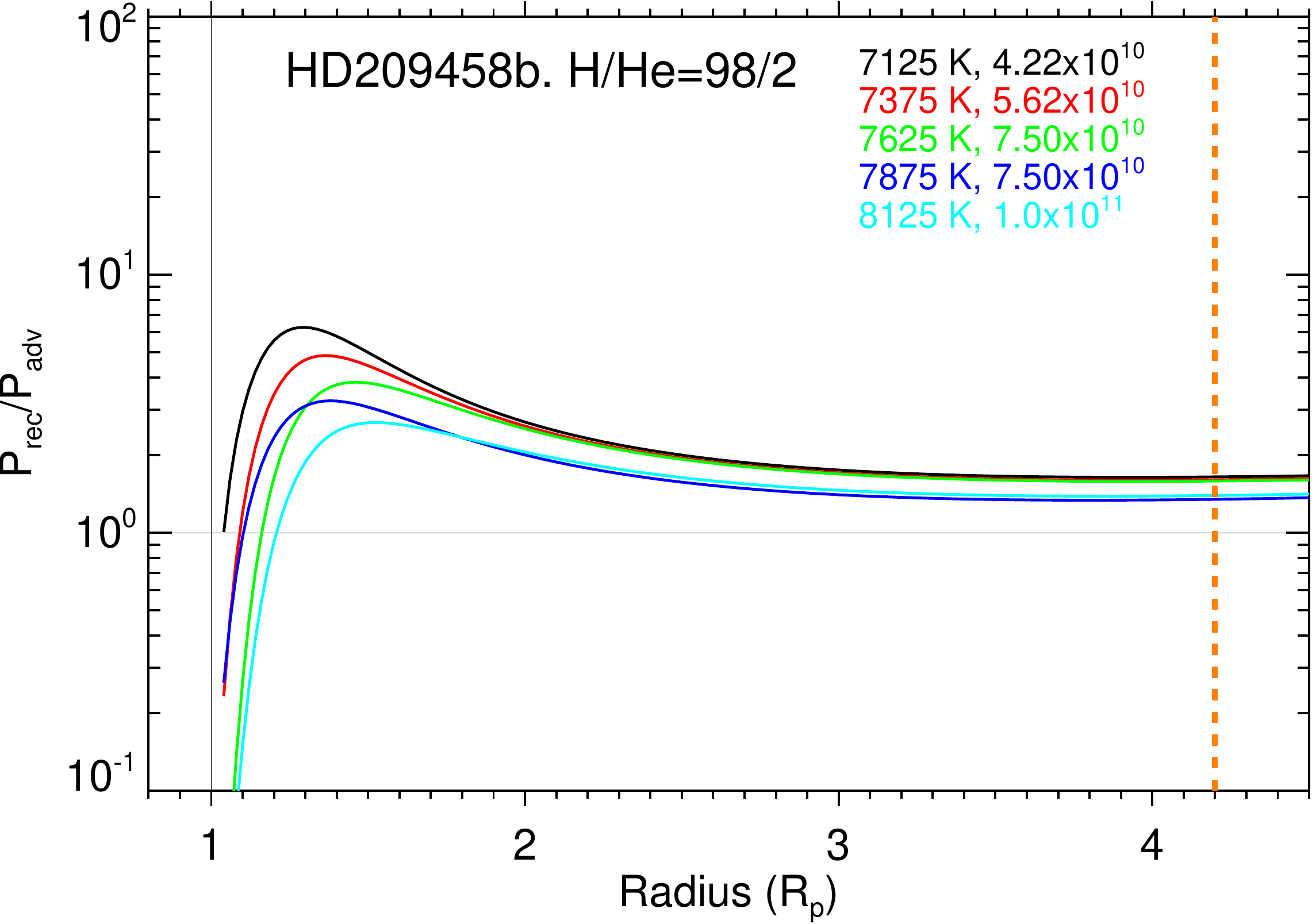}
\includegraphics[angle=0, width=0.66\columnwidth]{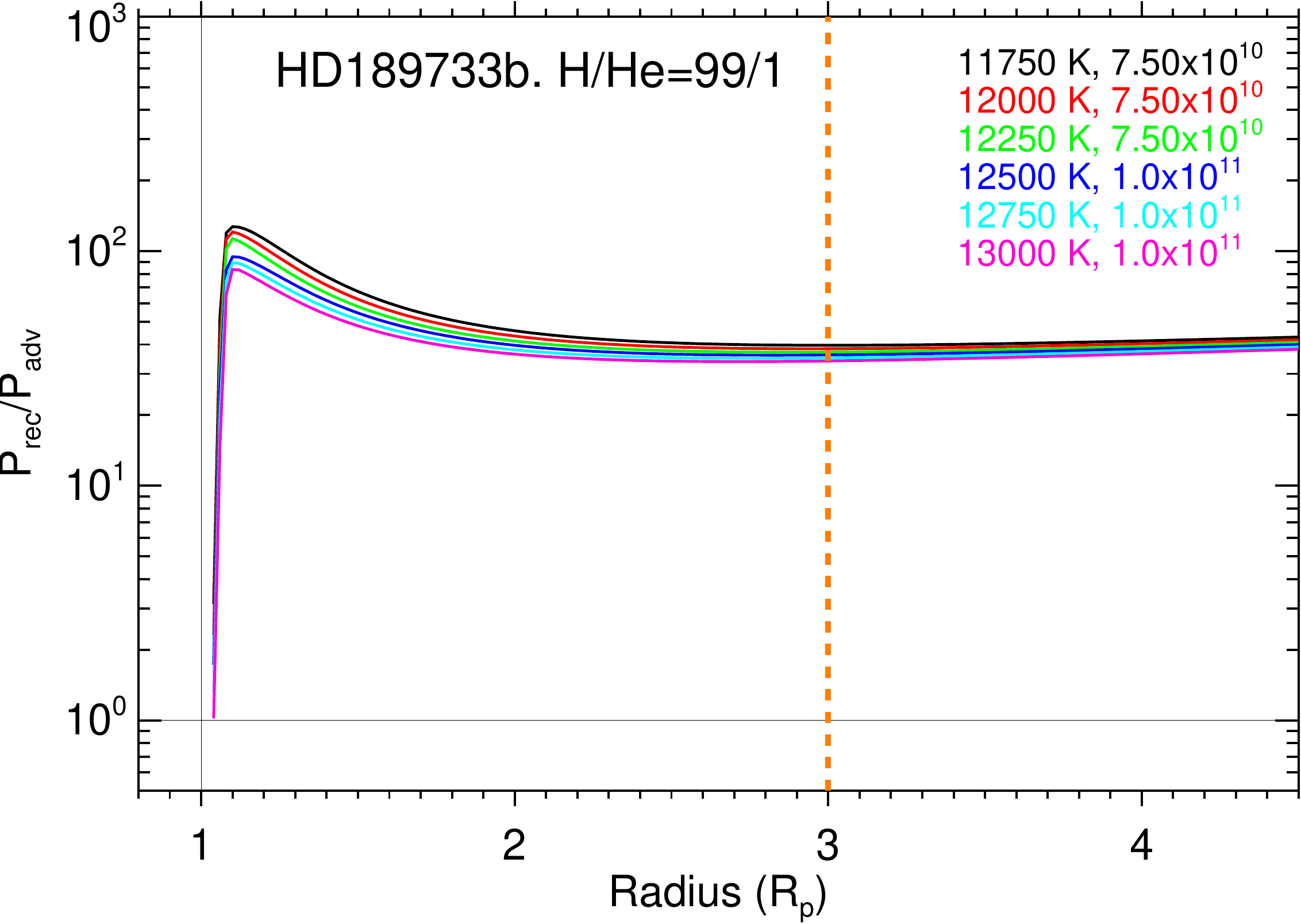}
\includegraphics[angle=0, width=0.66\columnwidth]{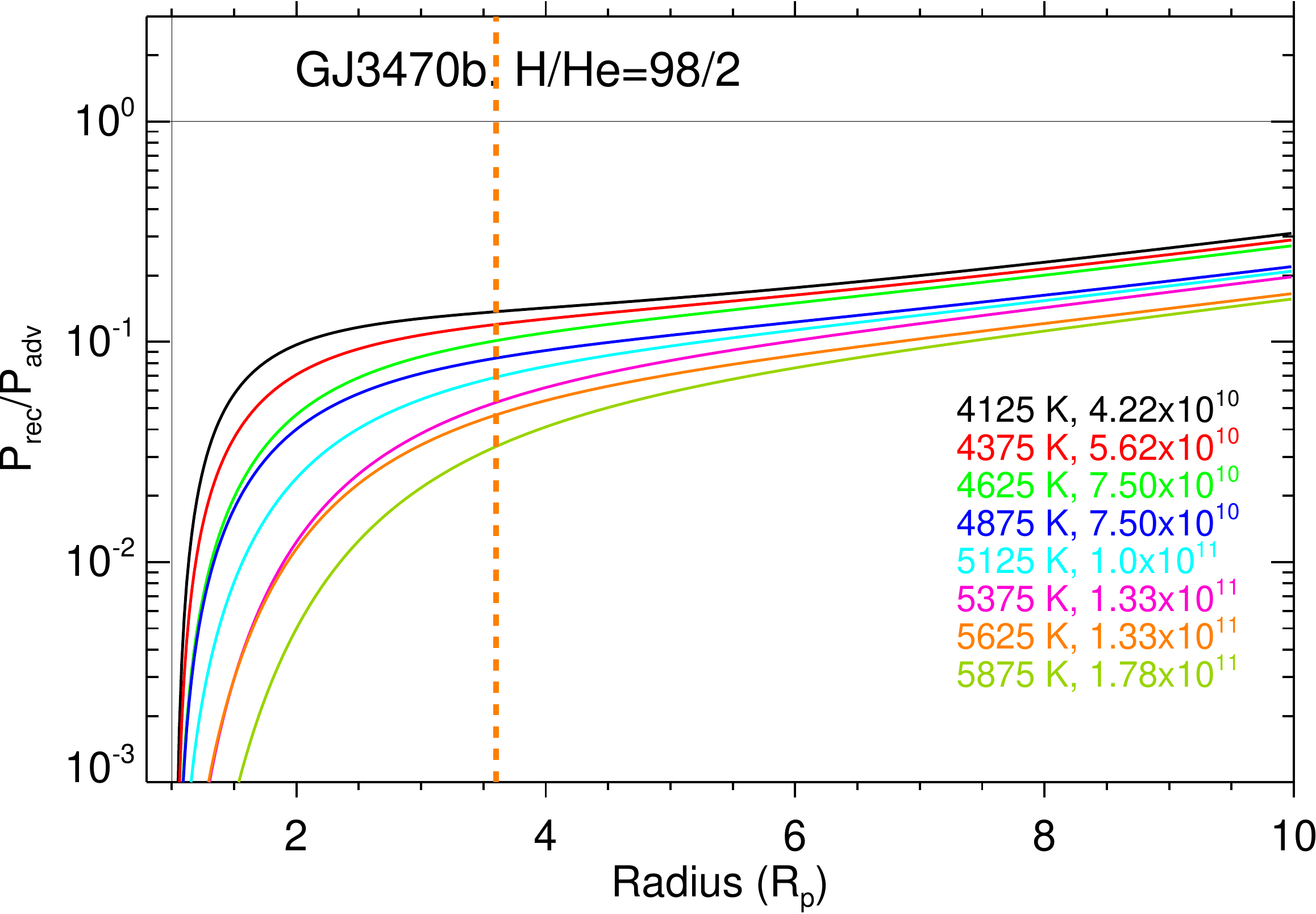}
\caption{
Recombination and advection for \hd20, \hdu18, and \gj34. Upper row: Production rates of H by recombination  (solid lines) and advection (dashed lines) derived from the fits of the measured \het\ absorption spectra (see symbols in Fig.\,\ref{chi2}) in the atmospheres of \hd20 (left), \hdu18 (middle), and \gj34 (right) and the representative H/He ratios of 98/2, 99/1, and 98/2, respectively. Lower row: Ratios of the recombination to the advection. The vertical dashed orange lines indicate the mean Roche lobes. The different scales of the x and y axes for the different planets should be noted.}
\label{hterms} 
\end{figure*}

\section{Results and discussion}
\label{analysis}

In this section we discuss the hydrodynamic escape regimes on the basis of the H$^0$ and H$^+$  distributions as well as on the basis of the heating efficiencies for the three planets.

\subsection{Structure, production, and losses of the neutral H}
\label{sec:ionization}

The H$^0$ mole fractions that result from the derived $T$--$\dot M$ ranges 
(see the symbols in Fig.\,\ref{chi2})
are shown in Fig.\,\ref{hmol}. 
The $f_{\rm {H^0}}$ profiles for \hdu18 show a very narrow IF; it is confined at 
the lower boundary (set at 1.02\,\rp\ for all three planets), well below its Roche lobe, which is located at 3.0\,\rp. Accordingly, this planet shows a wide, fully ionised region. 
In contrast, \gj34 shows a very extended IF, from about 1.2\,\rp\ to far beyond its Roche lobe at 3.6\,\rp. Therefore, no atmospheric region of \gj34 is fully ionised. 
The H$^0$ structure of \hd20 shows a relatively wide IF, extending from the lower boundary to about 2.3\,\rp, which is below its Roche lobe (at 4.2\,\rp); this IF is thus wider than that of \hdu18 and narrower than that of \gj34. Therefore, as in \hdu18,  there is a region that is almost fully ionised below the Roche lobe,  although it is narrower than the corresponding region in \hdu18.

Figure\,\ref{hterms} shows the ratio of recombination to advection terms, P$_{\textrm {rec}}$/P$_{\textrm {adv}}$, for the derived $T$--$\dot M$ range of
the measured \het\ and \lya\ absorption spectra for \hd20, \hdu18, and \gj34 (see Fig.\,\ref{chi2}). 
For the case of \hdu18, recombination dominates in practically the whole upper atmosphere.
In contrast, the  
H$^0$ density is dominated by advection in the whole upper atmosphere of \gj34.
In \hd20, the H$^0$ 
density is dominated by advection (P$_{\textrm {rec}}$/ P$_{\textrm {adv}}$\,$<$\,1)  in the lower part of the IF, but by recombination (P$_{\textrm {rec}}$/ P$_{\textrm {adv}}$\,$>$\,1) in the upper part of the IF  and above. That is, neither advection nor recombination rates are negligible in the upper atmosphere of \hd20.
Hence, according to the criteria discussed in Sect.\,\ref{sec:criteria_ratio}, these results show that \hdu18 is in the recombination-limited regime, \gj34 is in the photon-limited regime, and \hd20 is in the energy-limited regime.

We also analysed other plausible $T$--$\dot M$ values beyond the derived interval, hereafter the `extended' $T$--$\dot M$ range (see the dotted lines in Fig.\,\ref{chi2}). 
These $T$--$\dot M$ values are still compatible with the measured \het\ and \lya\ absorption; however, they are less likely as they yield a higher $\chi ^2$.  
With this, we aim to determine if they would affect the classification with respect to the escape regime.
Figures\,\ref{hterms_supp} and \ref{hmol_supp} show the results for the P$_{\textrm {rec}}$/P$_{\textrm {adv}}$ ratios and the corresponding H$^0$ mole fractions, respectively. 
We found that the extended $T$--$\dot M$ ranges 
for \hdu18\ and \gj34\ exhibit the same behaviour as that of the corresponding derived $T$--$\dot M$ ranges 
(see Fig.\,\ref{hterms}). In contrast, production terms for \hd20\ at temperatures below $\sim$\,7000\,K (see Fig.\,\ref{hterms_supp}) are at the limit between the energy-limited and recombination-limited regimes, with advection becoming less important relative to recombination.
However, when comparing the H density profiles for these conditions with those derived from  the \lya\ measurements by \cite{Garcia_munoz_2007},  \cite{Koskinen2013a}, and \cite{Salz2016}, we found that the H$^0$ density profiles are significantly smaller. 
We thus conclude that the hydrodynamic regimes  
discussed above for the three planets are not altered when considering the extended $T$--$\dot M$ ranges. 
We find that the classification of the escape regimes also remains unchanged in these exoplanets if the commonly used H/He fraction of 90/10 is assumed instead of the ratios derived from \het\ and \lya\ transmission spectroscopy (see Figs.\,\ref{hterms_supp2} and \ref{hmol_supp2}).

For \hdu18, we used an XUV flux that accounts for several small flares to provide an average model of active and non-active periods \cite[see Sect.\,3.2 in][]{Lampon2020b}. In order to explore the hydrodynamic regime of this planet in quiescent stages, we conducted a test using the XUV flux from the X-exoplanets model in \cite{Sanz_Forcada_2011}, which is about a factor of three smaller than the one used here. We found that the hydrodynamic regime remains recombination-limited despite such a considerable XUV flux decrease.

\subsection{Heating efficiencies}
\label{sec:efficiencies}

\begin{figure}[]
\includegraphics[angle=0, width=1.0\columnwidth]{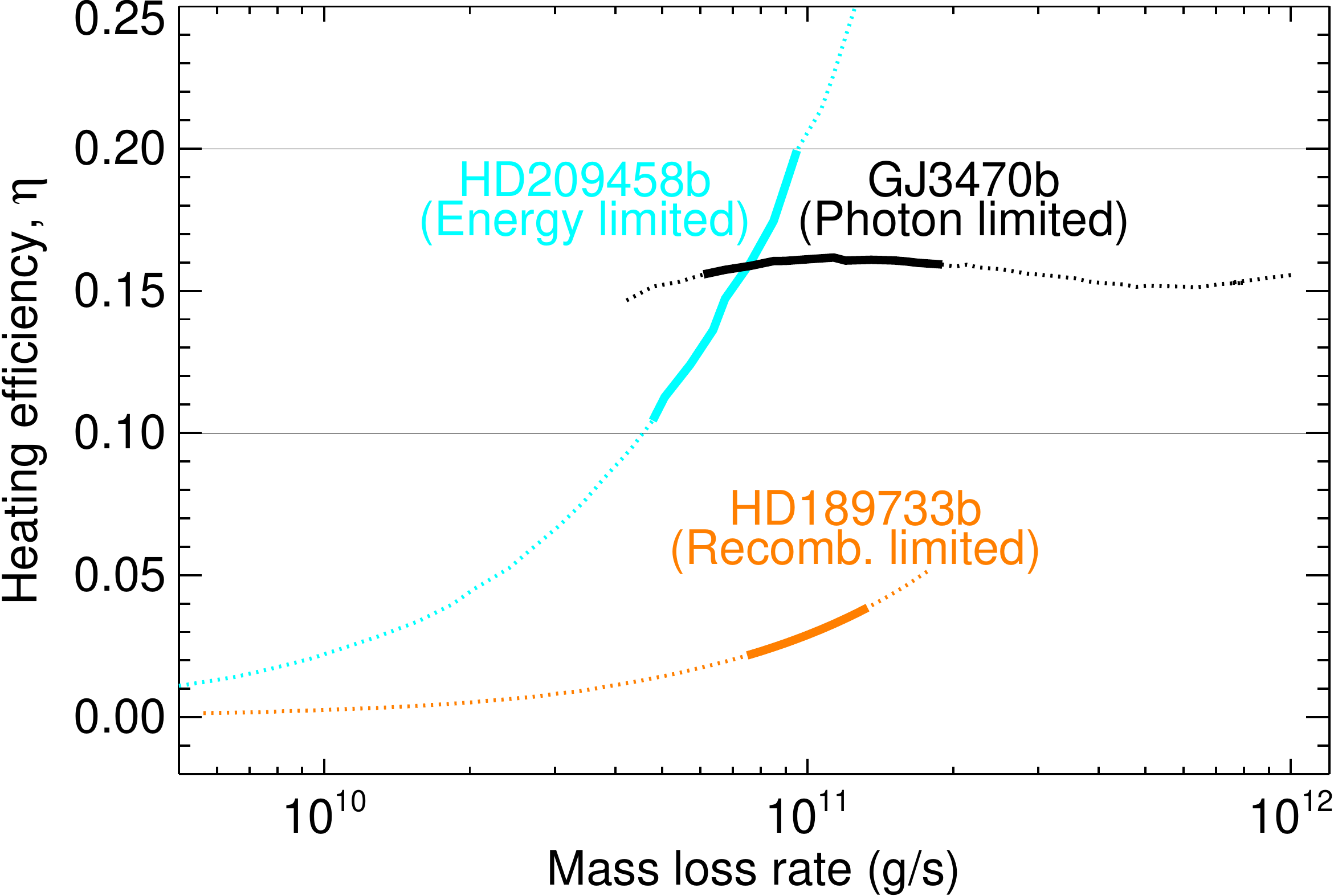} 
\caption{Heating efficiency, $\eta$, versus mass-loss rates for \hdu18 (orange), \hd20 (cyan), and \gj34 (black). H/He ratios of 99/1 for \hdu18 and of 98/2 for \hd20 and \gj34 were used. Dotted lines show the extended $T$--$\dot M$ ranges and thick lines the derived $T$--$\dot M$ ranges (see the dotted lines and symbols, respectively, in Fig.\,\ref{chi2}).}  
\label{eta}
\end{figure}

To calculate the heating efficiencies of the upper atmospheres of the planets, we followed the method from  \cite{Salz_2015}. Essentially, we used the relationship between the heating efficiency, $\eta$, and the mass-loss rate for the energy-limited approximation, $\dot M_{\rm EL}$ \citep{Watson_1981,Erkaev_2007}, together with the relationship \mlr/$\dot M_{\rm EL}$\,=\,4/5 derived by \cite{Salz_2015} to obtain
\begin{align}
\dot M = \frac{4}{5}\, \frac {4 \pi\, R_{\rm p}\, R_{\rm XUV}^{2}\,F_{\rm XUV}} {G\,K(\xi)\, M_{\rm P}}\, \eta, 
\label{eq:energy_lim}
\end{align}
where $M_{\rm p}$ and \rp \ are the planetary mass and radius, respectively; $F_{\rm XUV}$ is the XUV stellar flux at the planetary orbital separation; $R_{\rm XUV}$ is the effective absorption radius, the altitude where the XUV optical depth is unity; and $G$ is the gravitational constant. Here, $K(\xi)$\,=\,1--1.5\,$\xi+0.5\,\xi^{3}$ is the potential energy reduction factor, with $\xi$\,=\,$\left(M_{\rm P}/M_{\star} \right)^{1/3}\left(a/R_{\rm P}\right)$, where $a$ is the planetary orbital separation and $M_{\star}$ the stellar mass.
Therefore, taking the \mlr\ and $R_{\rm XUV}$ obtained from the \het\ measurements and the required system parameters from Table\,\ref{param}, we calculated their corresponding $\eta$ values (see Fig.\,\ref{eta}). The \mlr\ values are taken from Fig.\,\ref{chi2} (ranges limited by the symbols), and the $R_{\rm XUV}$ values are $\approx$1.02\,\rp\ for \hdu18\ (i.e. the lower boundary), about 1.16--1.30\,\rp\ for \hd20, and about 1.38--2.46\,\rp\ for \gj34.

As shown by the thick lines in Fig.\,\ref{eta}, $\eta$ reaches high values in \hd20 and \gj34 but remains very low ($\eta$\,$<$\,0.04) for \hdu18. 
This indicates that \hdu18 is in the recombination-limited regime, as predicted \cite[e.g. by][]{Salz_2015}.     

We note that the derived $T$--$\dot M$ range for \hd20\ was obtained by limiting the heating efficiency to the range of 0.1--0.2 \citep{Lampon2020a}, as shown in Fig.\,\ref{eta}.
Values of $\eta\leq$\,0.1 would indicate a recombination-limited regime (see Sect.\,\ref{sec:criteria_heating}). However, they correspond to temperatures below T\,$\approx$\,7000\,K, which, as discussed in Sect.\,\ref{sec:ionization}, were discarded on the basis of \lya\  observations. 
This as well as the heating efficiency argument led us to conclude that \hd20 is not in the recombination-limited regime.

Heating efficiency remains almost constant with $\dot M$ in \gj34, in contrast to \hd20. As the upper atmosphere of \gj34\ is highly extended, an increase in $\dot M$ implies higher $R_{\rm XUV}$ (i.e. a higher planetary XUV cross-section), which prevents $\eta$ from increasing. 
The fact that $\eta$ is nearly constant is equivalent to stating that $\dot M$\,$\propto$\,R$_{\rm XUV}^2$ (see Eq.\,\ref{eq:energy_lim}), which, according to \cite{Owen_2016}, indicates that \gj34\ is in the photon-limited regime.

\section{Summary} 
\label{summary}
Several theoretical studies have identified three distinct hydrodynamic atmospheric escape regimes
in hydrogen-dominated planetary atmospheres: energy-limited, recombination-limited, and photon-limited.
In addition to the importance from a theoretical point of view, the classification of the hydrodynamic escape regimes has important consequences for the estimation of the mass-loss rate by models based on the energy-limited approach.
However, no evidence of these regimes had been reported to date.

In this work, we studied the recombination to advection rates of the upper atmospheric H obtained in a previous analysis of the measured absorption spectrum of the He~{\sc i} triplet lines at 10830\,\AA\ and \lya\ of \hdu18, \hd20, and \gj34, which was performed by \cite{Lampon2020a,Lampon2020b}. Furthermore, based on these results, we also calculated the heating efficiencies, following \cite{Salz_2015}.

We found that in \hdu18 the neutral H production is governed by recombination in the whole upper atmosphere, giving rise to a very narrow IF.
In contrast, advection dominates the neutral H density in the entire upper atmosphere of \gj34, leading to a very extended IF.
\hd20 is an intermediate case between \hdu18 and \gj34: The advection and the recombination processes are non-negligible, which forms a relatively extended IF. 

The derived heating efficiencies for \hd20 and \gj34 are similar to those expected from heating via photoelectrons, while those of \hdu18 are much lower. Therefore, \hd20 and \gj34 show little loss by radiative cooling, in contrast to \hdu18. 
Additionally, we found that $\dot M$\,$\propto$\,R$_{\rm XUV}^2$ in \gj34, which leads to a nearly constant heating efficiency in the range of 0.15--0.17.
Following two criteria, one based on the productions and losses of the neutral H and the other based on 
energy balance considerations, we have shown that \hdu18 is in the recombination-limited regime, \gj34 is in the photon-limited regime, and \hd20 is in the energy-limited regime.

Our analysis involves a series of important model assumptions, such as
the use of 1D models without an explicit solution of the energy budget equation and assumed lower boundary conditions. We have shown that
these assumptions do not significantly affect the derived bulk properties,
such as thermospheric temperatures, mass-loss rates, radial velocities, H/He ratios, and, in particular, the recombination and advection rates. Consequently, we are confident that our analysis of measured \het\ and \lya\ absorption provides robust evidence for the existence of the three theoretically predicted hydrodynamic escape regimes in hydrogen-dominated atmospheres.
Accordingly, \hd20, \hdu18, and \gj34 can be considered as benchmark cases for their respective regimes.

\begin{acknowledgements}
We thank the referee for very useful comments.
CARMENES is an instrument for the Centro Astron\'omico Hispano-Alem\'an (CAHA) at Calar Alto (Almer\'{\i}a, Spain), operated jointly by the Junta de Andaluc\'ia and the Instituto de Astrof\'isica de Andaluc\'ia (CSIC).
CARMENES was funded by the Max-Planck-Gesellschaft (MPG), the Consejo Superior de Investigaciones Cient\'{\i}ficas (CSIC), the Ministerio de Econom\'ia y Competitividad (MINECO) and the European Regional Development Fund (ERDF) through projects FICTS-2011-02, ICTS-2017-07-CAHA-4, and CAHA16-CE-3978, and the members of the CARMENES Consortium (Max-Planck-Institut f\"ur Astronomie,
Instituto de Astrof\'{\i}sica de Andaluc\'{\i}a,
Landessternwarte K\"onigstuhl, 
Institut de Ci\`encies de l'Espai, 
Institut f\"ur Astrophysik G\"ottingen, 
Universidad Complutense de Madrid, 
Th\"uringer Landessternwarte Tautenburg, 
Instituto de Astrof\'{\i}sica de Canarias, 
Hamburger Sternwarte, 
Centro de Astrobiolog\'{\i}a and 
Centro Astron\'omico Hispano-Alem\'an), 
with additional contributions by the MINECO, the Deutsche Forschungsgemeinschaft through the Major Research Instrumentation Programme and Research Unit FOR2544 ``Blue Planets around Red Stars'', the Klaus Tschira Stiftung, the states of Baden-W\"urttemberg and Niedersachsen, and by the Junta de Andaluc\'{\i}a.
We acknowledge financial support from the Agencia Estatal de Investigaci\'on of the Ministerio de Ciencia, Innovaci\'on y Universidades and the ERDF  through projects
ESP2016--76076--R,
ESP2017--87143--R,
BES--2015--074542,
PID2019-110689RB-I00/AEI/10.13039/501100011033, 
PGC2018-099425--B--I00,
PID2019-109522GB-C51/2/3/4,     
PGC2018-098153-B-C33,           
AYA2016-79425-C3-1/2/3-P,       
ESP2016-80435-C2-1-R,           
and the Centre of Excellence ``Severo Ochoa'' and ``Mar\'ia de Maeztu'' awards to the Instituto de Astrof\'isica de Canarias (SEV-2015-0548), Instituto de Astrof\'isica de Andaluc\'ia (SEV-2017-0709), and Centro de Astrobiolog\'ia (MDM-2017-0737), and the Generalitat de Catalunya/CERCA programme.
T.H. acknowledges support from the European Research Council under the Horizon 2020 Framework Program via the ERC Advanced Grant Origins 832428.
A.S.L. acknowledges funding from the European Research Council under the European Union's Horizon 2020 research and innovation program under grant agreement No 694513.
\end{acknowledgements}

\bibliographystyle{aa} 
\bibliography{ref.bib}

\begin{appendix}

\section{The steady-state continuity equation of neutral H}
\label{ap:derivation}

The hydrodynamic continuity equation of mass can be expressed as
\begin{equation}
    \dot M = 4\, \pi \,r^2\, \rho_{\rm r} \, v_{\rm r},
    \label{eq:hidrod:1}
\end{equation}
where $\rho_{\rm r} = n_{\rm H}\,m_{\rm H}\,+\,n_{\rm He}\,m_{\rm He}$; $n_{\rm H}$ and $n_{\rm He}$ are the number densities of H and He, respectively; and $m_{\rm H}$ and $m_{\rm He}$ are the atomic masses of H and He, respectively.
Assuming that $n_{\rm H}/n_{\rm He}$ (hereafter H/He) is constant with respect to $r$, then $n_{\rm He} =  (H/He)^{-1}\,n_{\rm H}$. 
Therefore, $\rho_{\rm r} =  n_{\rm H}\,x $,  where $x = m_{\rm H}\,+\,(H/He)^{-1}\,m_{\rm He}$, is constant. As such, the mass-loss rate can be expressed as 
\begin{equation}
    \dot M = 4\, \pi \,r^2\, n_{\rm H}\, x \, v_{\rm r}. 
    \label{eq:hidrod:2}
\end{equation}
The continuity equation of H$^0$ can be written as
\begin{equation}
    \frac{1}{r^2} \frac{\partial ( r^2 \,n_{\rm H^0}\,v_{\rm r}) }{\partial r}  = n_{\rm H^+}\,n_{\rm e}\,\alpha_{\rm H} -  n_{\rm H^0}\,J_{\rm H}
    \label{eq:cont_H:1}
\end{equation}
and, deriving, as\begin{equation}
    \frac{2\,n_{\rm H^0}\,v_{\rm r}}{r}\,+\,v_{\rm r}\,\frac{\partial n_{\rm H^0}}{\partial r}\,+\,n_{\rm H^0}\frac{\partial v_{\rm r}}{\partial r}  = 
    n_{\rm H^+}\,n_{\rm e}\,\alpha_{\rm H} -  n_{\rm H^0}\,J_{\rm H}.
    \label{eq:cont_H:2}
\end{equation}
Taking $ v_{\rm r} = \dot M /(4\, \pi \,r^2\, n_{\rm H}\, x) $ from Eq.\, \ref{eq:hidrod:2} and re-arranging Eq.\ A.4, we have
\begin{equation}
        v_{\rm r}\,\frac{1}{n_{\rm H}}\, \left[ n_{\rm H}\, \frac{\partial n_{\rm H^0}}{\partial r} - n_{\rm H^0}\,\frac{\partial n_{\rm H}}{\partial r} \right]= 
        n_{\rm H^+}\,n_{\rm e}\,\alpha_{\rm H} -  n_{\rm H^0}\,J_{\rm H}.
    \label{eq:cont_H:3}
\end{equation}
Taking into account that
\begin{equation}
        \frac{\partial f_{\rm H^0}}{\partial r} = \frac{\partial \left(  n_{\rm H^0}/n_{\rm H} \right )}{\partial r} = \frac{1}{{n_{\rm H}^2}} \left ( \,n_{\rm H}\,\frac{\partial n_{\rm H^0}}{\partial r}\,-\,n_{\rm H^0}\,\frac{\partial n_{\rm H}}{\partial r} \right )
\end{equation}
and including it in Eq.\,\ref{eq:cont_H:3}, we obtain
\begin{equation}
        v_{\rm r}\,n_{\rm H}\,\frac{\partial f_{\rm H^0}}{\partial r}=
        n_{\rm H^+}\,n_{\rm e}\,\alpha_{\rm H} -  n_{\rm H^0}\,J_{\rm H}.
    \label{eq:cont_H:4}
\end{equation}
Assuming $n_{\rm e}$\,=\,$n_{\rm H^+}$, and re-arranging Eq.\,\ref{eq:cont_H:4}, we finally obtain Eq.\,\ref{eq:H}.

\section{Other tables and figures}\label{ap:table}

\begin{figure*}[htbp]
\includegraphics[angle=0, width=0.66\columnwidth]{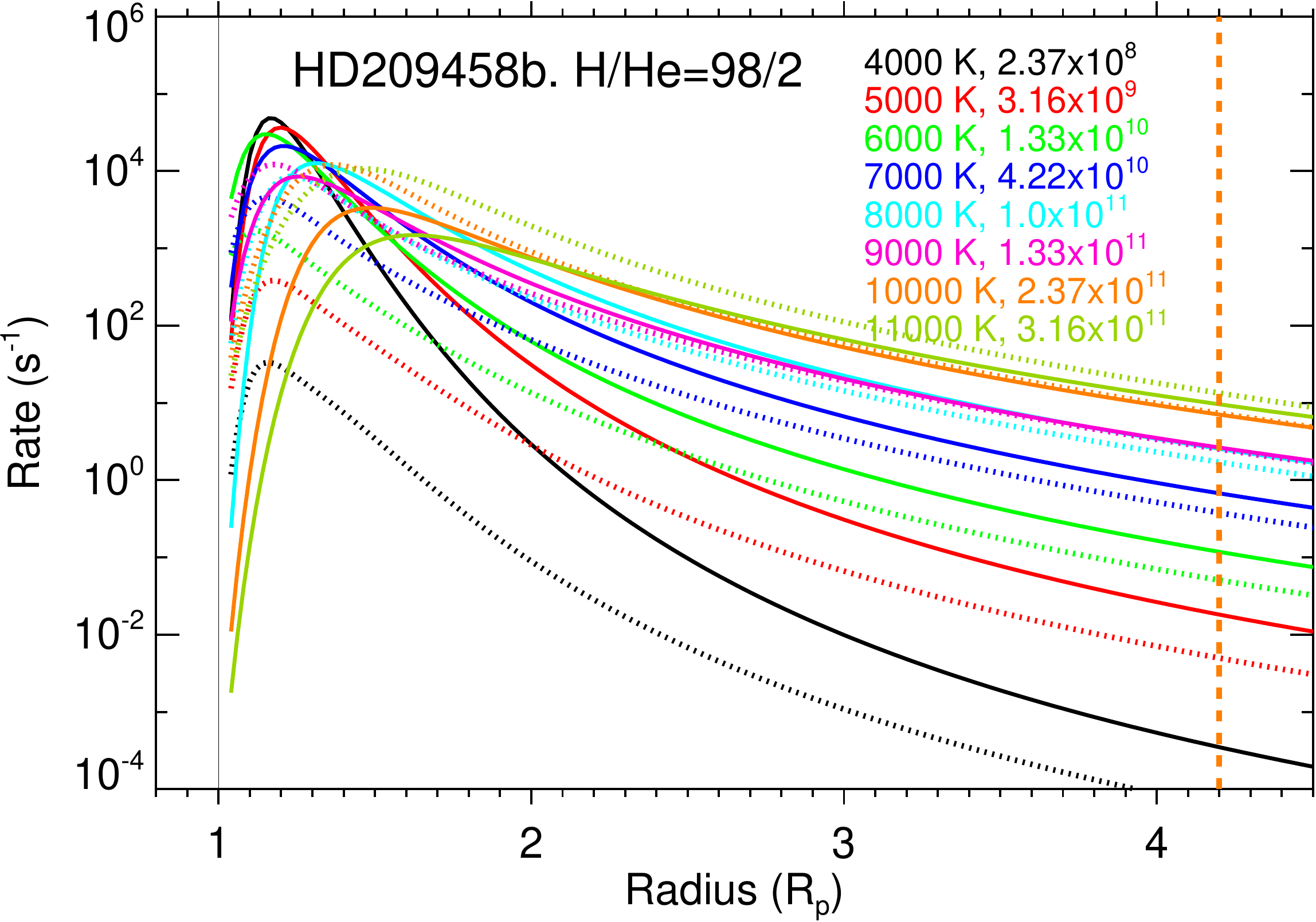}
\includegraphics[angle=0, width=0.66\columnwidth]{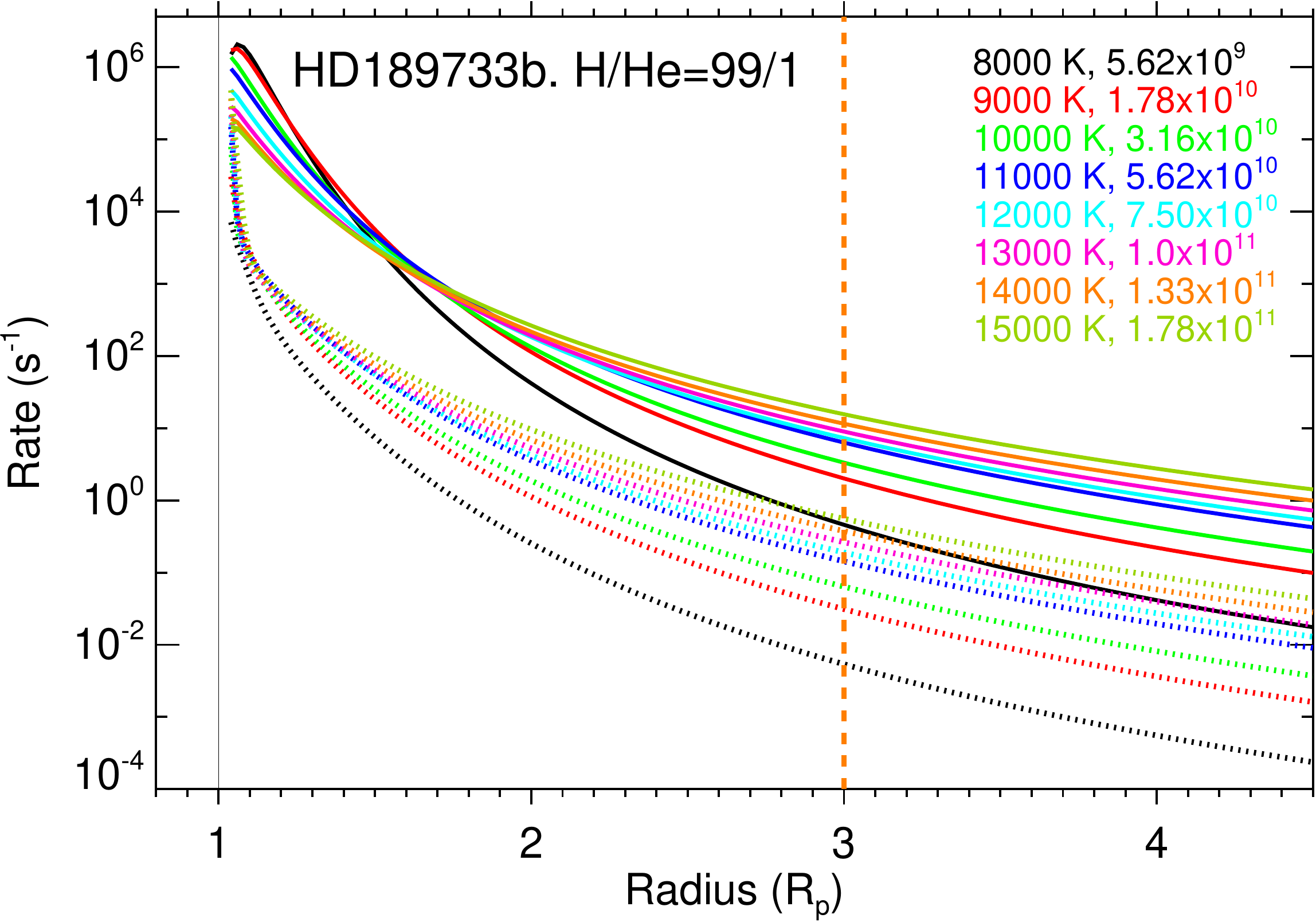}
\includegraphics[angle=0, width=0.66\columnwidth]{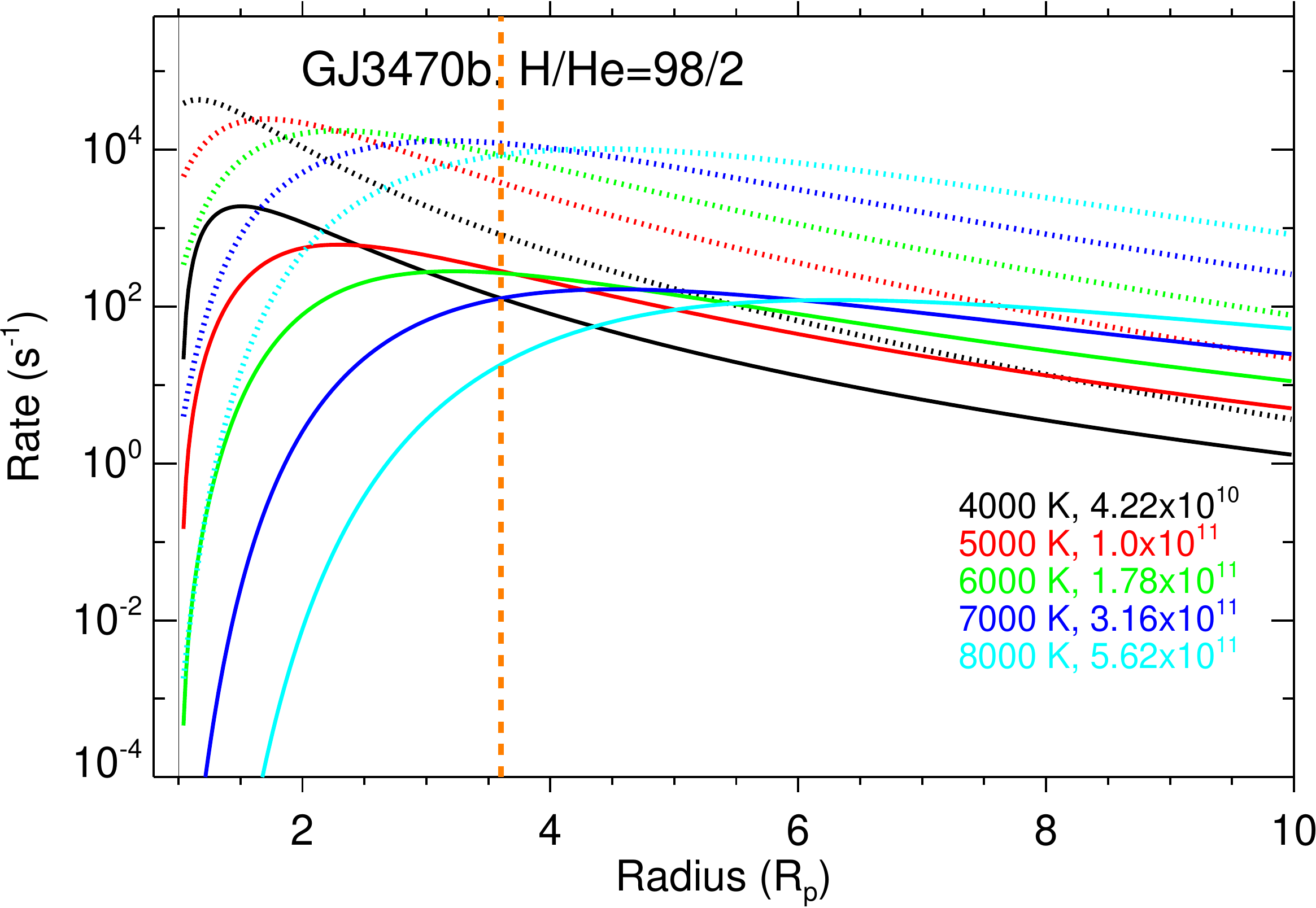}

\includegraphics[angle=0, width=0.66\columnwidth]{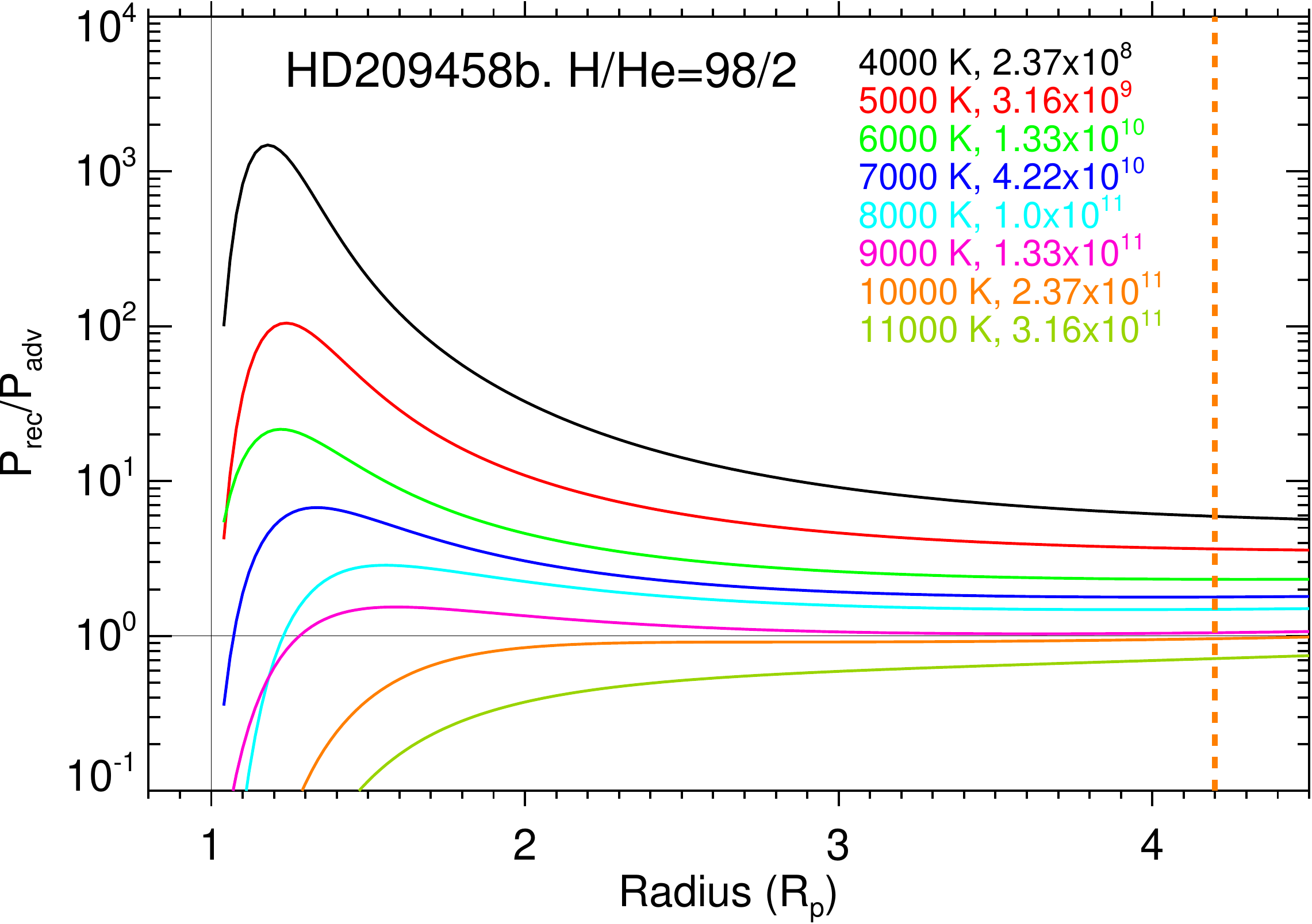}
\includegraphics[angle=0, width=0.66\columnwidth]{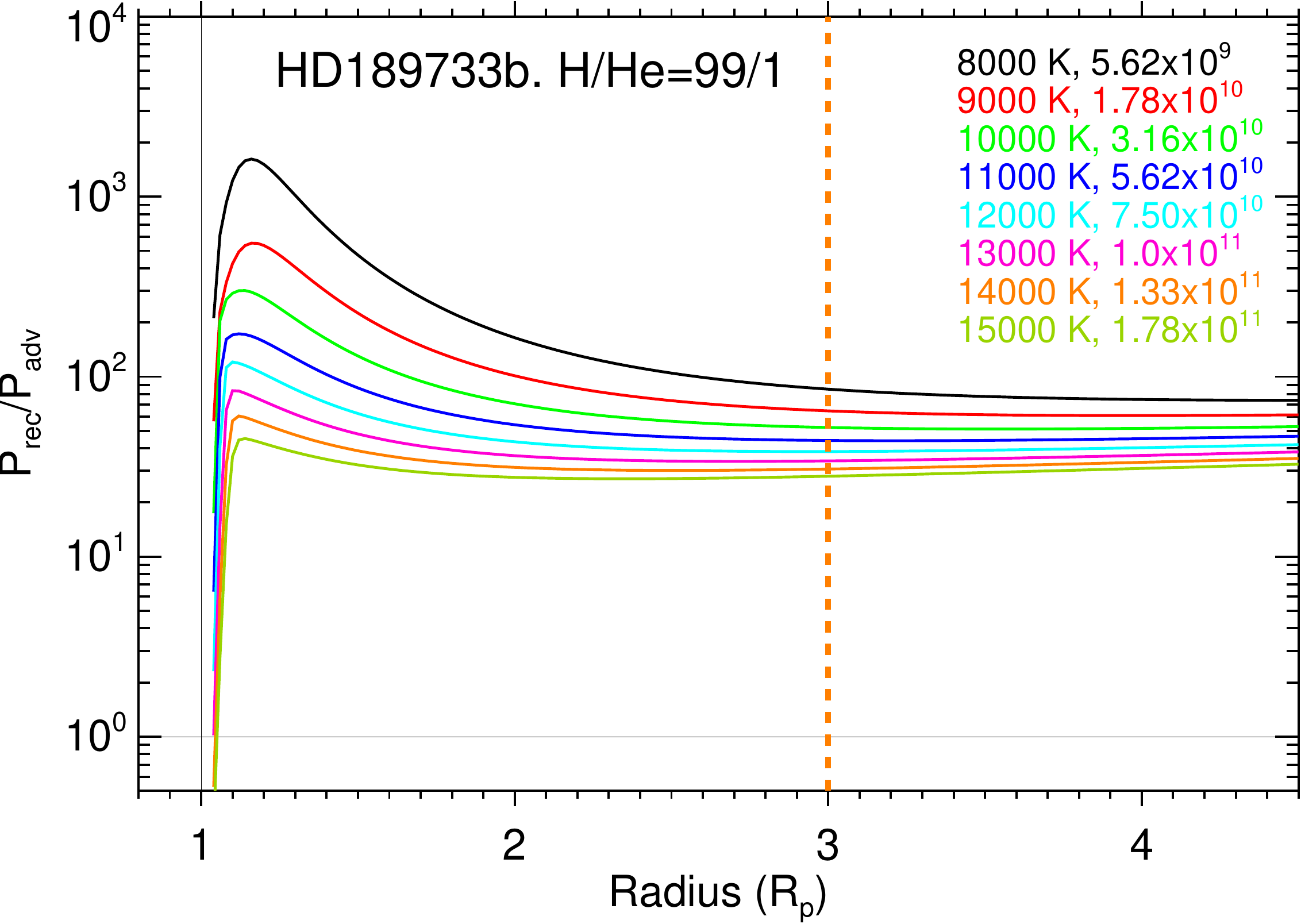}
\includegraphics[angle=0, width=0.66\columnwidth]{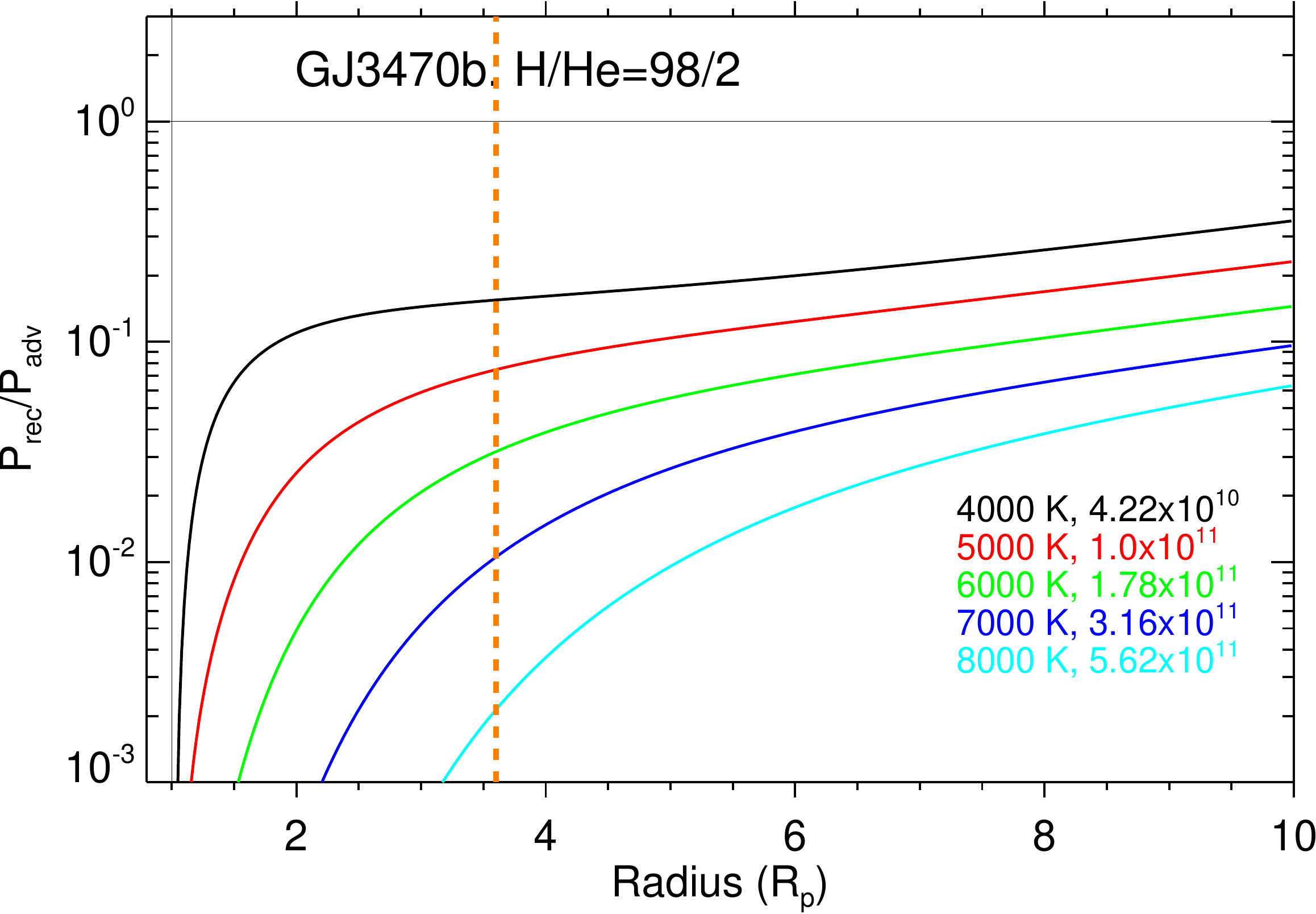}
\caption{Same as Fig.\,\ref{hterms} but for the extended $T$--$\dot M$ ranges (see the dotted lines in Fig.\,\ref{chi2}). Upper row: Rates of H by recombination (solid lines) and advection (dotted lines). Lower row: Ratios of recombination to advection.  The vertical dashed orange lines indicate the mean Roche lobes. We note the different scales of the x and y axes.}
\label{hterms_supp} 
\end{figure*}

\begin{figure*}[htbp]
\includegraphics[angle=0, width=0.66\columnwidth]{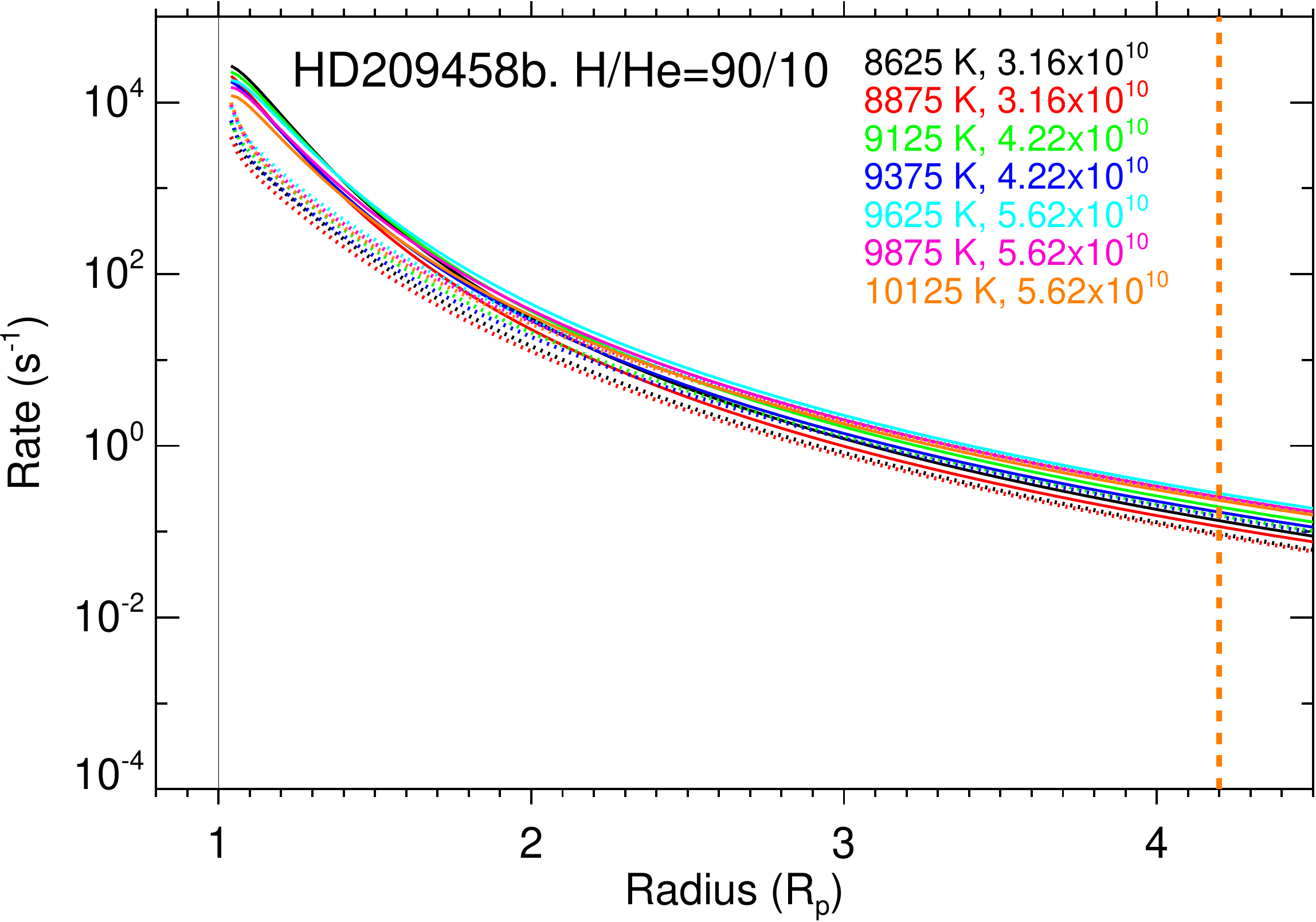}
\includegraphics[angle=0, width=0.66\columnwidth]{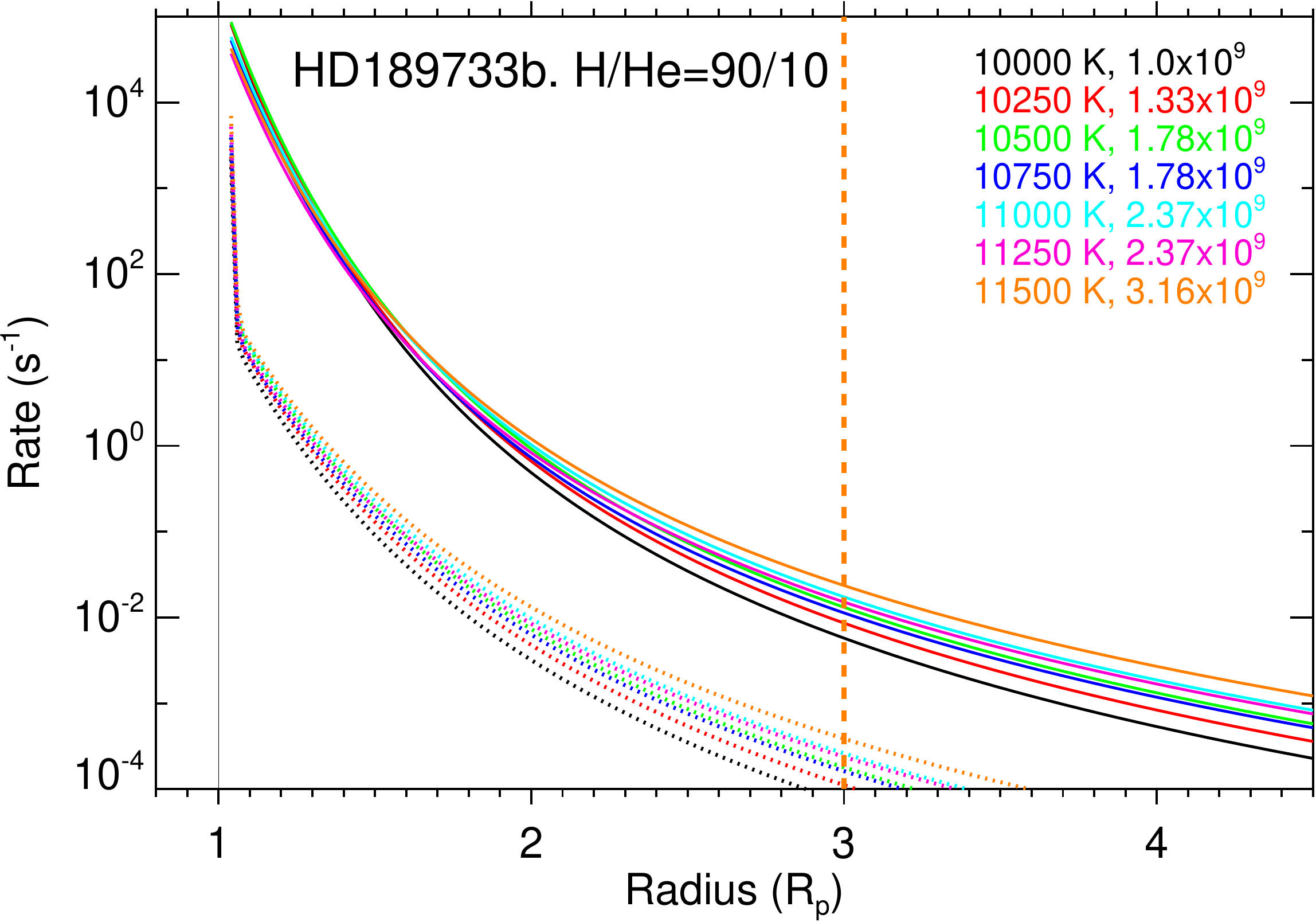}
\includegraphics[angle=0, width=0.66\columnwidth]{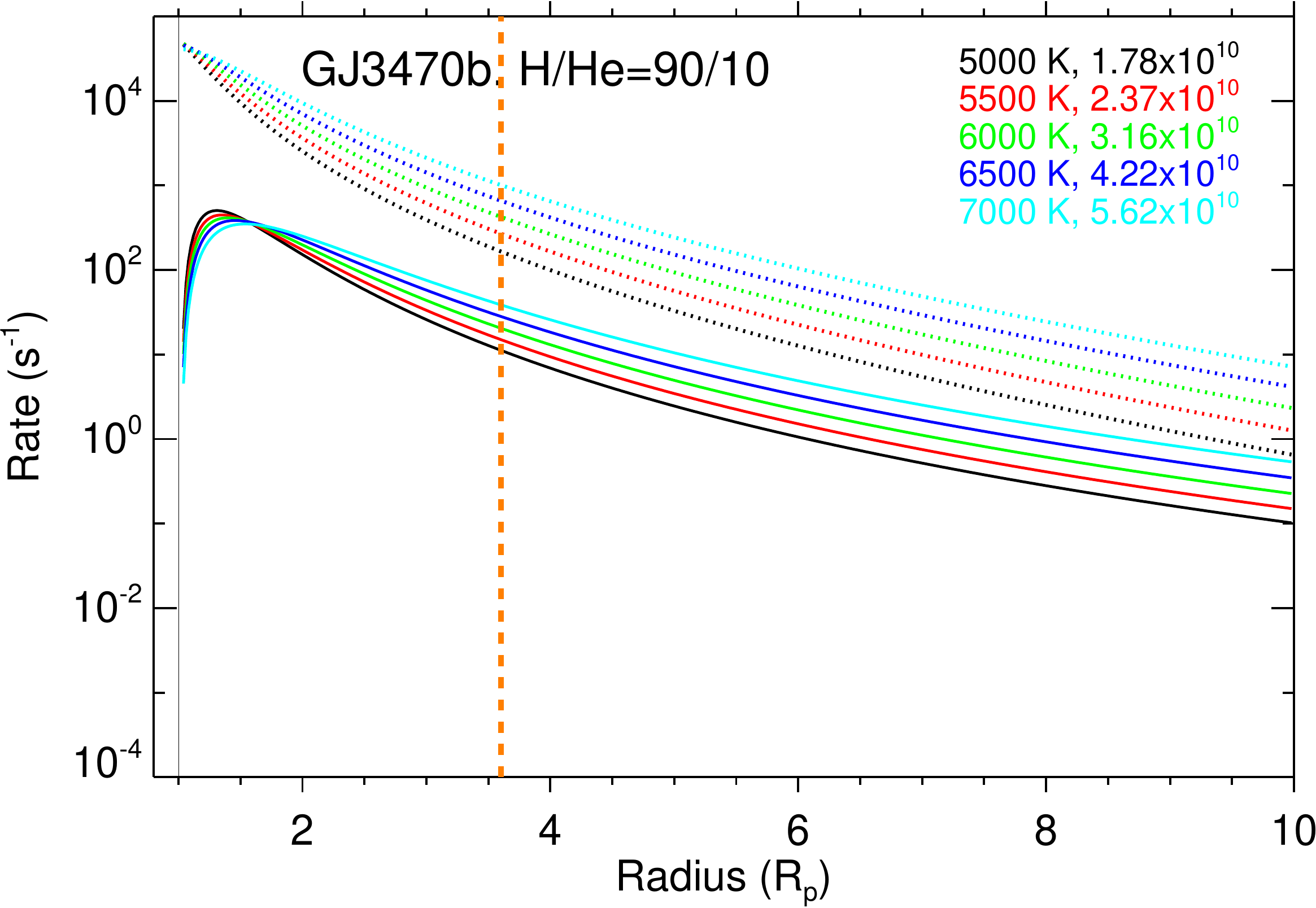}

\includegraphics[angle=0, width=0.66\columnwidth]{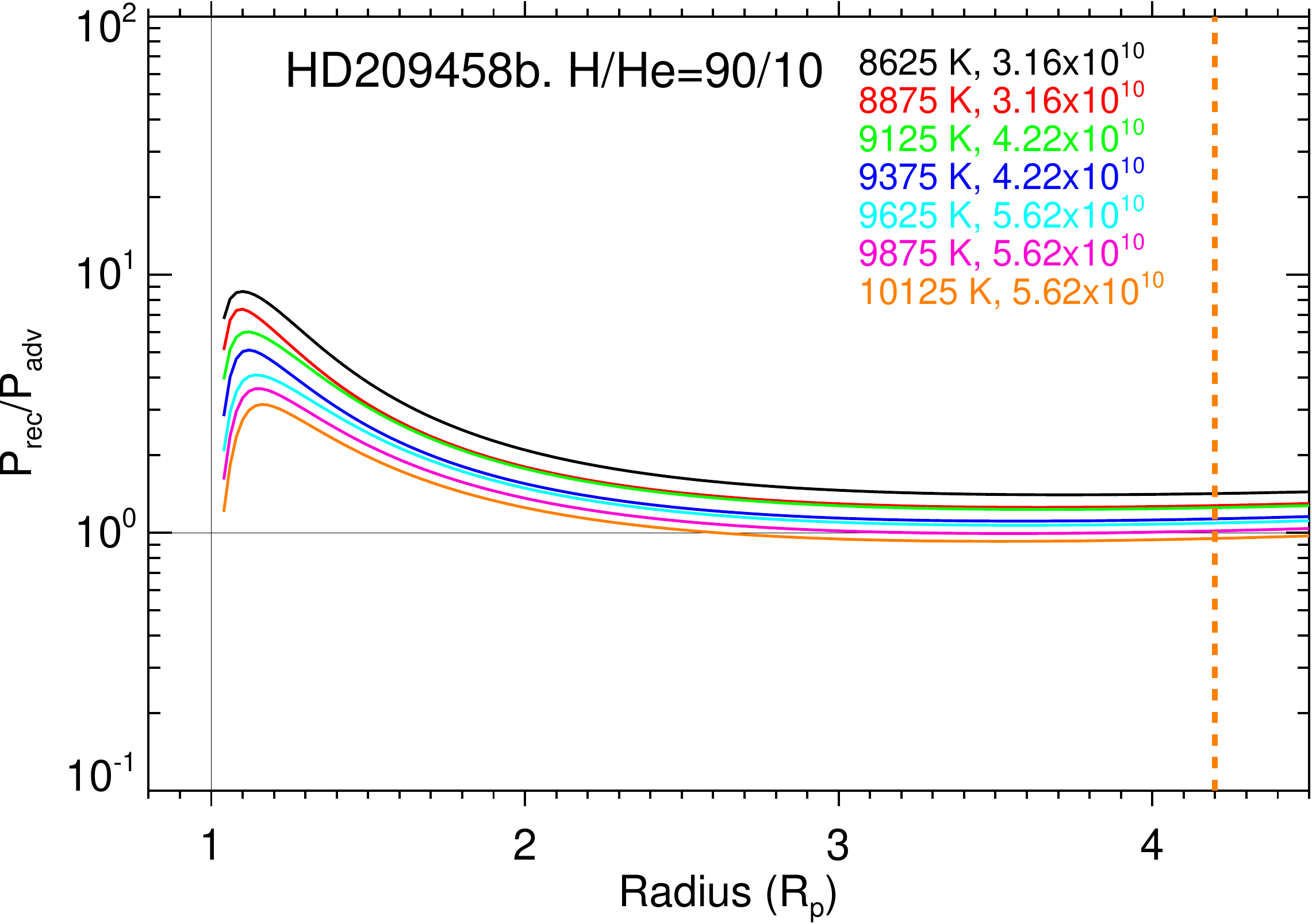}
\includegraphics[angle=0, width=0.66\columnwidth]{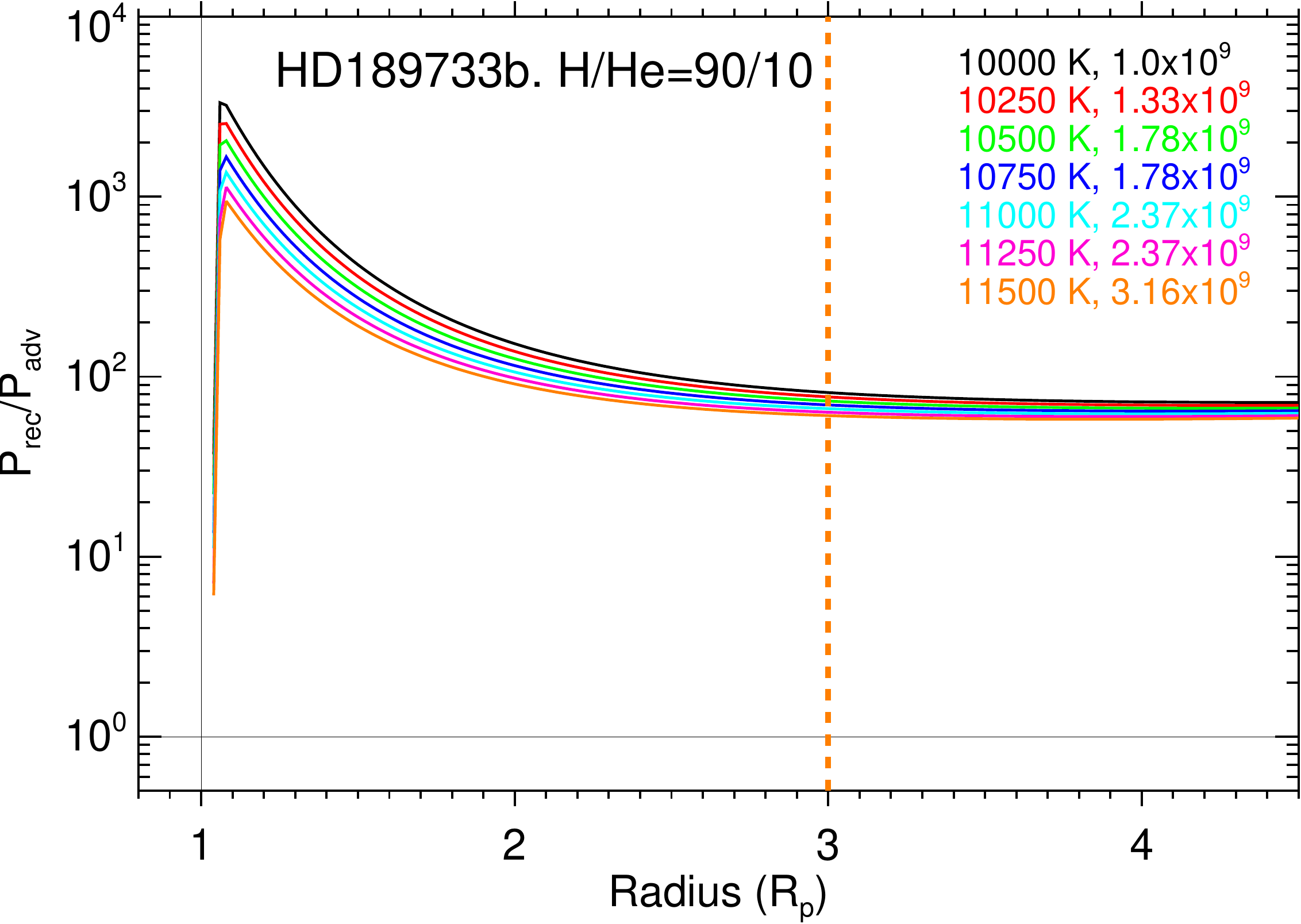}
\includegraphics[angle=0, width=0.66\columnwidth]{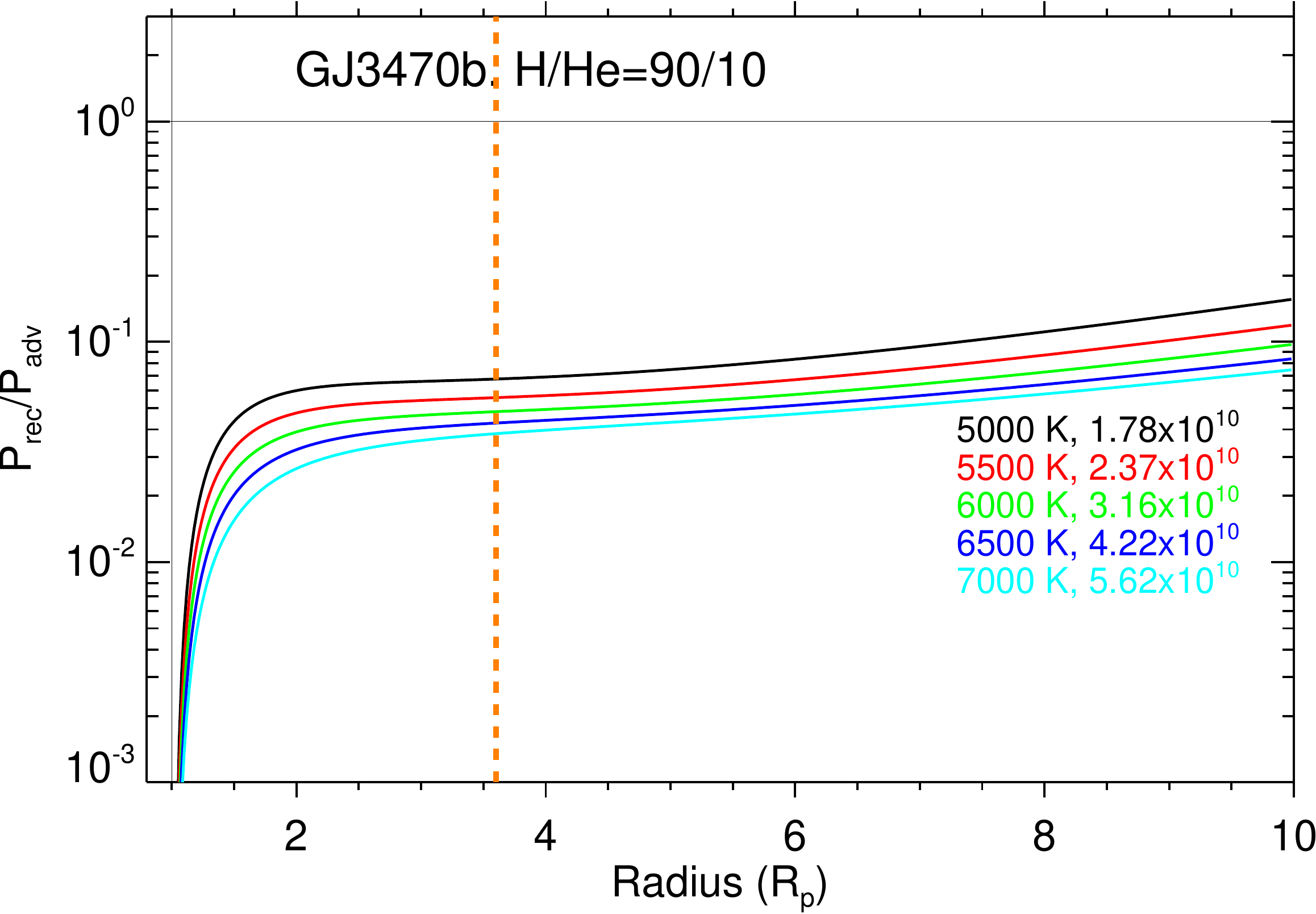}
\caption{Same as Fig.\,\ref{hterms} but for an H/He ratio of 90/10 \citep[see Fig. 5 in][]{Lampon2020b}. Upper row: Rates of H by recombination (solid lines) and advection (dotted lines). Lower row: Ratios of recombination to advection. The vertical dashed orange lines indicate the mean Roche lobes. We note the different scales of the x and y axes.}
\label{hterms_supp2} 
\end{figure*}

\begin{figure}[htbp]
\includegraphics[angle=0, width=\columnwidth]{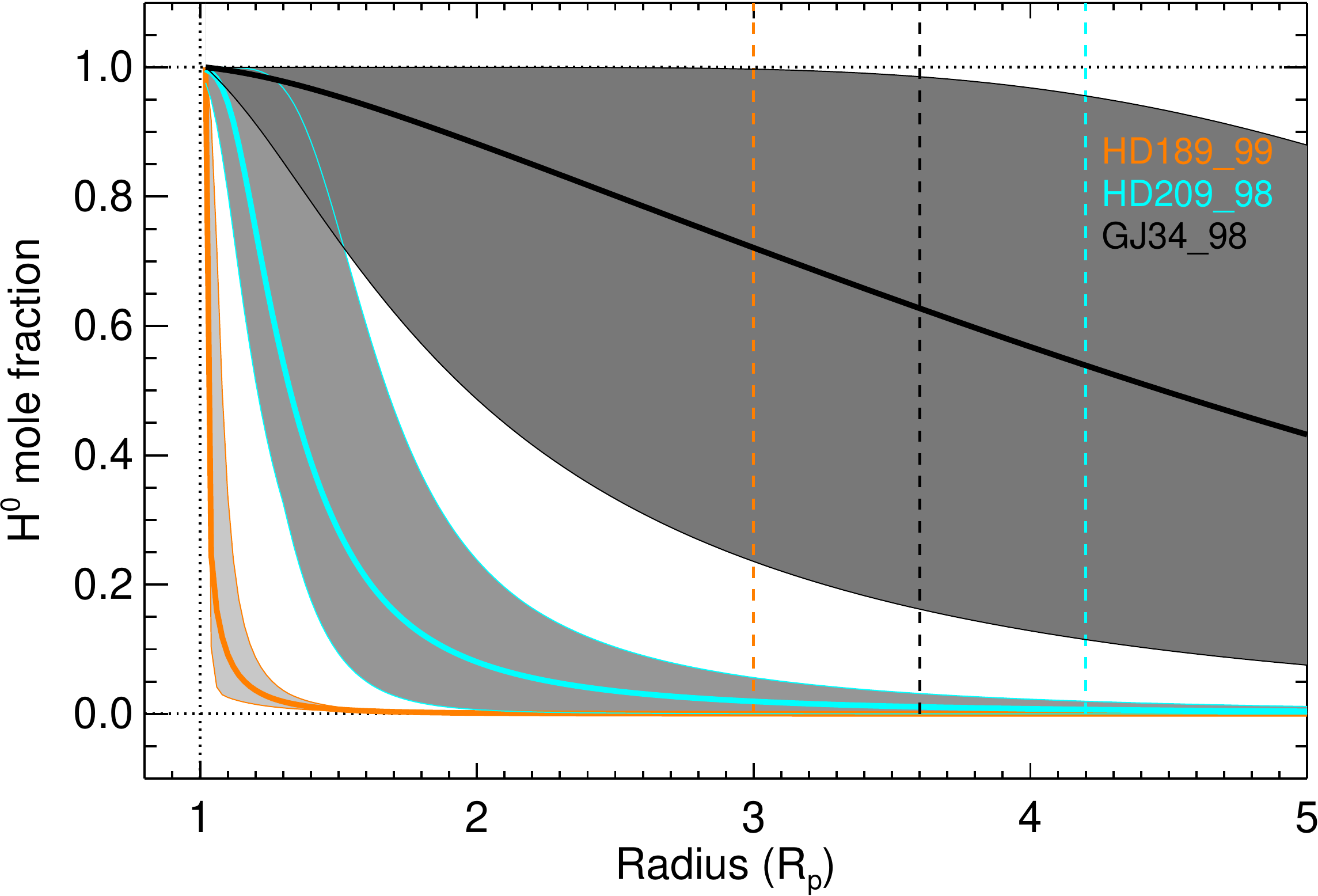}
\caption{Same as Fig.\,\ref{hmol} but for the extended $T$--$\dot M$ ranges (see the dotted lines in Fig.\,\ref{chi2}). The thicker solid lines are the mean profiles. The vertical dashed lines are the mean Roche lobes.}
\label{hmol_supp} 
\end{figure}

\begin{figure}[htbp]
\includegraphics[angle=0, width=\columnwidth]{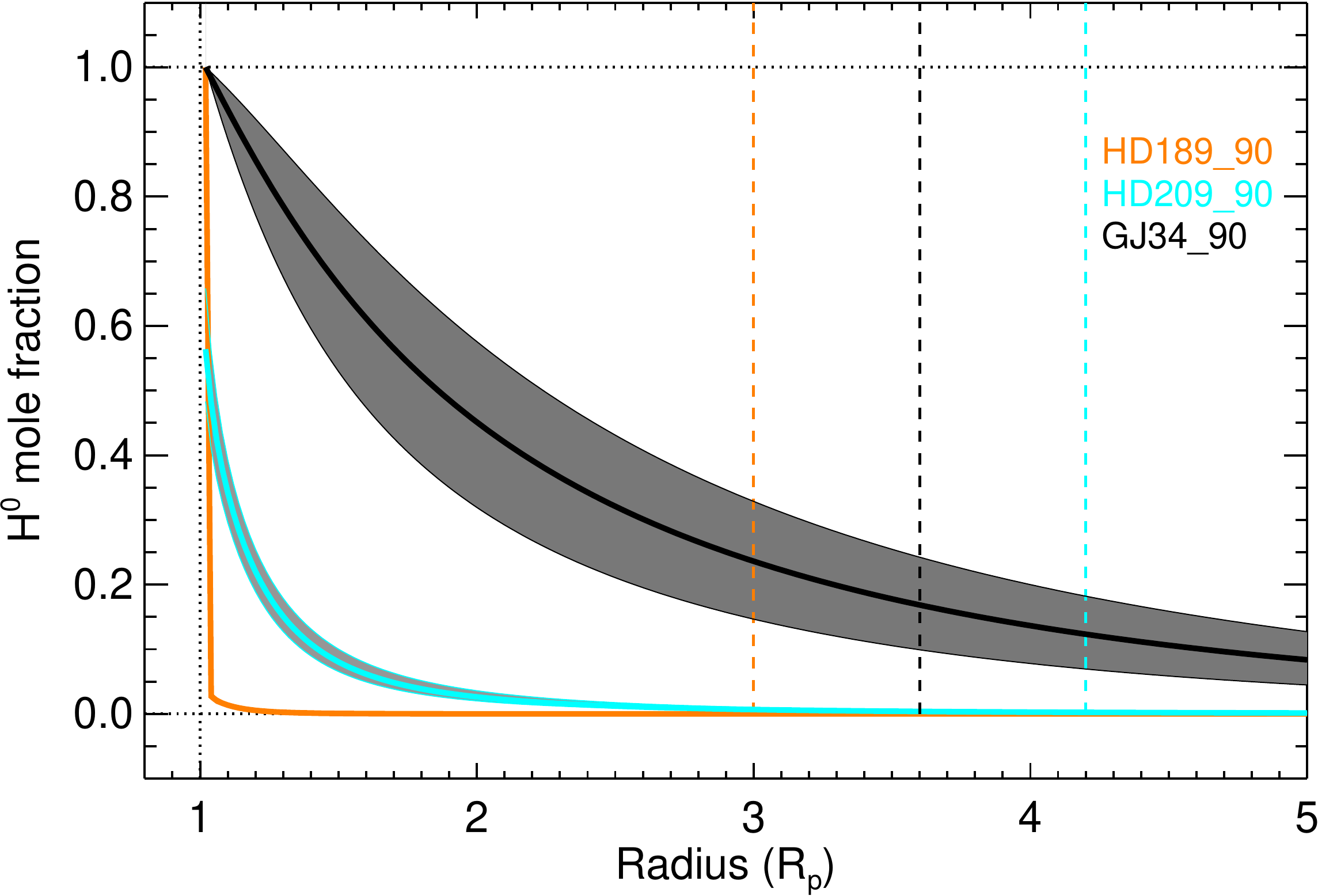}
\caption{Same as Fig.\,\ref{hmol} but for an H/He mole fraction of 90/10 \citep[see Fig. 5 in][]{Lampon2020b}. The thicker solid lines are the mean profiles. The vertical dashed lines are the mean Roche lobes.}
\label{hmol_supp2} 
\end{figure}

In Table\,\ref{param} we provide the most relevant system parameters used in our calculations. In Figs.\,\ref{hterms_supp} to \ref{hmol_supp2} we show ratios of recombination to advection and H$^0$ mole fraction profiles for the extended $T$--$\dot M$ ranges (dotted lines in Fig.\,\ref{chi2}) as well as for the H/He composition of 90/10 \citep[see the symbols in Fig.\,5 of][]{Lampon2020b}.

\begin{table}[htbp]
\centering
\caption{\label{param}System parameters.}
\begin{tabular}{l c c c } 
\hline  \hline  \noalign{\smallskip}
System                  & HD 209458 & HD 189733  & GJ 3470 \\
\hline \noalign{\smallskip}
M$_\star$\,(M$_{\odot}$)\,$^{(a)}$      & 1.119$^{+0.033}_{-0.033}$ & 0.846$^{+0.06}_{-0.049}$ & 0.476$^{+0.019}_{-0.019}$ \\
\noalign{\smallskip}
$a$\,(au)\,$^{(b)}$     & 0.04707$^{+0.00046}_{-0.00047}$ & 0.0332$^{+0.001}_{-0.001}$ & 0.0348$^{+0.0014}_{-0.0014}$ \\
\noalign{\smallskip}
M$_\textrm{P}$ (M$_{\textrm{Jup}}$)\,$^{(c)}$   & 0.685$^{+0.015}_{-0.014}$ & 1.162$^{+0.058}_{-0.039}$ & 0.036$^{+0.002}_{-0.002}$ \\
\noalign{\smallskip}
\rp\,(R$_{\textrm{Jup}}$)\,$^{(c)}$     & 1.359$^{+0.016}_{-0.019}$     & 1.23$^{+0.03}_{-0.03}$ &
0.360$^{+0.01}_{-0.01}$\\
\noalign{\smallskip}
$\Phi$\,($\Phi_{\rm Jup}$)\,$^{(d)}$    & 0.504 & 0.944 &  0.100 \\
\noalign{\smallskip}
R$_{\textrm{lobe}}$ (\rp)\,$^{(e)}$     & 4.2 &  3.0 &  3.6 \\
\noalign{\smallskip}
F$_{XUV}$\,$^{(f)}$ & 2.358     &  56.740       &  3.928 \\
\noalign{\smallskip}
\hline
\end{tabular}
\tablefoot{
\tablefoottext{a}{Stellar masses of HD 209458, HD 189733, and GJ 3470 from \cite{Torres08}, \cite{de_Kok_2013}, and \cite{Palle2020}, respectively.}
\tablefoottext{b}{Orbital separation from \cite{Torres08},  \cite{Agol_2010}, and \cite{Bonfils_2012}. }
\tablefoottext{c}{
Planetary mass and radius of \hd20\  from \cite{Torres08}; of \hdu18\ from \cite{de_Kok_2013} and \cite{Baluev_2015}; and of \gj34 from \cite{Palle2020}. }
\tablefoottext{d}{Gravitational potential.}
\tablefoottext{e}{Roche lobes of \hd20, \hdu18, and \gj34 by \cite{Salz2016}, \cite{Eggleton_1983}, and \cite{Bourrier2018}, respectively. }
\tablefoottext{f}{XUV flux in units of 10$^{3}$ erg\,cm$^{-2}$\,s$^{-1}$ at $\lambda$ < 912 \AA\ at the planetary separation calculated by \cite{Lampon2020a} for \hd20 and by \cite{Lampon2020b} for \hdu18 and \gj34.} 
}
\end{table}

\end{appendix}

\end{document}